\documentclass[a4paper,11pt]{article}
\pdfoutput=1

\usepackage{jcappub}

\usepackage[T1]{fontenc} 

\usepackage{ulem} 

\usepackage{comment}

\usepackage{amssymb}
\usepackage{amsmath}

\usepackage{array,booktabs}
\newcolumntype{P}[1]{>{\centering\arraybackslash}p{#1}}

\title{\boldmath Morphological Analysis of the Polarized Synchrotron Emission with WMAP and Planck}

\author[a]{F. A. Martire,}
\author[b]{A. J. Banday,}
\author[a]{E. Mart\'{\i}nez-Gonz\'alez}
\author[a]{and R. B. Barreiro}

\affiliation[a]{Instituto de F\'isica de Cantabria, CSIC-Universidad de Cantabria,\\ 
Avda. de los Castros s/n, E-39005 Santander, Spain}
\affiliation[b]{IRAP, Université de Toulouse, CNRS, CNES, UPS, Toulouse, France}

\emailAdd{martire@ifca.unican.es}

\abstract{
The bright polarized synchrotron emission, away from the Galactic plane, originates mostly from filamentary structures. We implement a filament finder algorithm which allows the detection of bright elongated structures in polarized intensity maps. We analyse the sky at 23 and 30~GHz as observed respectively by \textit{WMAP} and \textit{Planck}. We identify 19 filaments, 13 of which have been previously observed. For each filament, we study the polarization fraction, finding values typically larger than for the areas outside the filaments, excluding the Galactic plane, and a fraction of about 30\% is reached in two filaments. We study the polarization spectral indices of the filaments, and find a spectral index consistent with the values found in previous analysis (about –3.1) for more diffuse regions. Decomposing the polarization signals into the $E$ and $B$ families, we find that most of the filaments are detected in $P_E$, but not in $P_B$. We then focus on understanding the statistical properties of the diffuse regions of the synchrotron emission at 23~GHz. Using Minkowski functionals and tensors, we analyse the non-Gaussianity and statistical isotropy of the polarized intensity maps. For a sky coverage corresponding to 80\% of the fainter emission, and on scales smaller than 6 degrees ($\ell > 30$), the deviations from Gaussianity and isotropy are significantly higher than 3$\sigma$. The level of deviation decreases for smaller scales, however, it remains significantly high for the lowest analised scale ($\sim 1.5^\circ$).  When 60\% sky coverage is analysed, we find that the deviations never exceed 3$\sigma$. Finally, we present a simple data-driven model to generate non-Gaussian and anisotropic simulations of the synchrotron polarized emission. The simulations are fitted in order to match the spectral and statistical properties of the faintest 80\% sky coverage of the data maps.
}

\begin{document}
\maketitle
\flushbottom


\section{Introduction}
\label{sec:01}
The principle focus of studies of the Cosmic Microwave Background (CMB) in recent years has been, and indeed remains, to detect, and subsequently characterize, its polarized emission. Precision measurements are necessary in order to detect the very weak $B$ modes generated by primordial gravitational waves, as specifically predicted by models of inflation. However, the sky emission at radio/microwave wavelengths also contains various foreground sources of astrophysical emission that completely obscure the cosmological signal. Specifically, at frequencies below a few GHz, the emission is dominated by synchrotron and free–free radiation; above 10~GHz the so-called Anomalous Microwave Emission (AME) contribution becomes significant; and above 70~GHz the thermal dust emission becomes dominant. Synchrotron and dust radiations are highly polarized, the free-free radiation is intrinsically unpolarized, and AME is expected to be only weakly polarized \cite{G_nova_Santos_2016}. 

Synchrotron radiation is due to relativistic cosmic ray (CR) electrons accelerating around the Galactic magnetic field. The emission intensity depends on the density and energy distributions of the electrons, and on the Galactic magnetic field strength. The electron energy distribution can be approximated by a power-law, $N(E) \propto E^{-p}$ \cite{adriani2011cosmic, ackermann2012fermi}, with typical values close to $p = 3$ \cite{ackermann2010fermi}. In a uniform magnetic field, the synchrotron radiation can reach a polarization fractional of $\Pi=(p + 1)/(p + 7/3) \approx 75\%$. However, we observe much smaller values, because of a geometric depolarization due to tangled magnetic fields and superposition effects along the line-of-sight. In addition, at frequencies typically below 10~GHz, Faraday rotation effects can depolarize the synchrotron emission near the Galactic plane \cite{FuskelandSynch}. 

The synchrotron spectral index is related to $p$ via $\beta = -(p + 3)/2$. Typical values around -3 have been reported for spectral index both in intensity \cite{davies2006determination,ade2015joint} and polarization \cite{akrami2020planck,NicolettaSych, Martire_2022}. More recent analyses seem to indicate the presence of spatial variations in the spectral index \cite{Svalheim2020BeyondPlanckXP,2022MNRAS.517.2855D, QJ_Elena}, with a tendency to steeper values moving from low to high Galactic latitudes \cite{FuskelandSynch}. 

The all-sky map at 408~MHz \cite{haslam1981408, haslam1982408, remazeilles2015improved}, often referred to hereafter as the Haslam map, provides the best full-sky representation of the synchrotron intensity emission. This is mainly due to cosmic rays accelerated by shock fronts in supernova remnants (SNRs) and pulsar wind nebulae (PWN). Outside the Galactic plane, the strong emission originates mostly from filamentary structures. The North Polar Spur (NPS), or Loop I, is the most obvious feature, but others have been observed: the Cetus arc (or Loop II) \cite{Large62}, Loop III \cite{Quigley65} and Loop IV \cite{Large66}. Those filaments are even more visible in the polarized sky. \cite{Vidal_2015} identify and study 11 filaments in the \textit{WMAP} polarization maps \cite{Bennett_2013}. The true origins of filaments are still poorly understood. The most widely accepted progenitors of these large structures are old and nearby supernova remnants \cite{spoelstra1973galactic}. 

The presence of complex structures such as loops and filaments makes the statistics of the synchrotron emission strongly non-Gaussian and anisotropic at large scales, even in the diffuse region. However, it is reasonable to suppose that at small scales the emission could approach Gaussianity and isotropy as a manifestation of the central limit theorem. Several models used to simulate synchrotron assume that the small scale fluctuations are statistically isotropic and Gaussian \cite{Tegmark_2000, Jelic08, PySM,waelkens2009simulating}. Analyzing the 408~MHz map, \cite{Rahman21} showed that the level of the non-Gaussian deviations decreases on smaller scales as expected, but remains significantly high ($>3\sigma$) on angular scales of $\sim 1.5^\circ$. These results were confirmed by analyzing WMAP and Planck intensity maps \cite{Rahman_22}.

However, little is known in polarization. Knowing the morphological and statistical properties of the polarized foregrounds emission is crucial to face the future challenges detecting cosmological signals. Some component separation methods used to produce CMB maps require prior knowledge of the foregrounds \cite{CompSep2016,akrami2020planck}. Moreover, the increasing sensitivity of on-going and future experiments, such as the Simons Observatory \cite{Ade_2019} and \textit{LiteBIRD} \cite{Matsumura_2014}, requires more realistic foreground models and simulations. From another perspective, characterizing the synchrotron emission can give us useful information for understanding the physical mechanisms behind the Galactic magnetic field \cite{Vidal_2015}. 

In this work, we characterize some morphological and statistical features of the synchrotron polarization. We analyze the observations of the \textit{WMAP} K-band and \textit{Planck} 30~GHz frequency channels in a region of the sky where the emission is predominantly diffuse. Section~\ref{sec:02} describes the data set and simulations used for the analysis. In section~\ref{sec:03}, we present a filament finder algorithm and demonstrate its performance on data maps. Section~\ref{sec:04} contains the description and analysis of the polarization fractions, spectral indices, $E$ and $B$ nature, and possible intensity counterparts of the detected filaments. In section~\ref{sec:05}, we test the statistical properties of the  \textit{WMAP} polarization maps. Section~\ref{sec:06} presents a simple model to generate simulations which better resemble the statistical nature of the polarized synchrotron. We summarize our results and provide discussion about their implications in section~\ref{sec:07}. Finally, in appendix~\ref{sec:a01} we give a brief review of some cosmological quantities, in appendix~\ref{sec:a02} we test the accuracy and limits of the finder algorithm, and in appendix~\ref{sec:a03} we present the results obtained from the \textit{Planck} statistical analysis.

\section{Polarized Intensity}
\label{sec:02}

\subsection{Data}
\label{sec:02,sub:01}
For our analysis, we will make use of data taken by the \textit{WMAP} and \textit{Planck} satellites. We focus on the lowest frequency data from \textit{WMAP}, specifically the 9-year \textit{WMAP} \textit{K}-band (centered at 23~GHz) maps, provided in the \texttt{HEALPix}\footnote{https://healpix.sourceforge.io} pixelisation scheme with $N_{side} = 512$ and an effective Gaussian beam of 0.88$^\circ$ full-width-at-half-maximum (FWHM). The \textit{WMAP} products have been downloaded from the Legacy Archive for Microwave Background Data Analysis (LAMBDA)\footnote{lambda.gsfc.nasa.gov/product/map}. For the \textit{Planck} analysis, we use the 30~GHz frequency maps generated by the \texttt{NPIPE} processing pipeline (PR4). The \texttt{NPIPE} processing results in improved High Frequency Instrument (HFI) polarization data with reduced systematic artefacts and lower levels of noise. PR4 data from the Low Frequency Instrument (LFI) are also modified with respect to the 2018 \textit{Planck} release. Further details are available in \cite{Npipe}. The frequency maps were downloaded from the \textit{Planck} Legacy Archive\footnote{pla.esac.esa.int} (PLA) at a pixel resolution corresponding to $N_{side} = 1024$ and an effective beam of FWHM = 31.5 arcminutes. Note that the polarization maps, and consequentially the analysis, follow the \texttt{HEALPix} convention.

As the synchrotron emission scales with frequency, the foreground signal is higher in the \textit{WMAP} K-band compared to the \textit{Planck} 30~GHz channel, however, the noise level of \textit{Planck} is lower. As result, at a common resolution of $1^\circ$, the overall signal-to-noise ratio of the two experiments are similar \cite{Planck2016Diff}, although one or other map may be better in some sky regions because of the different scanning strategies. 

\subsection{Smoothing}
\label{sec:02,sub:02}
We smooth the \textit{WMAP} and \textit{Planck} maps to common resolutions of $1^\circ$ and $3^\circ$ FWHM. The $1^\circ$ maps are downgraded to a \texttt{HEALPix} resolution of $N_{side} = 128$ (corresponding to a representative pixel size of $\sim$27 arcmin) that we will use for filament detection and statistical analysis, and the $3^\circ$ maps to a resolution of $N_{side} = 64$ (a pixel size of $\sim$55 arcmin), that we will use for the filament analysis. The smoothing and downgrading are performed in harmonic space deconvolving the original effective beam and then convolving with a Gaussian beam\footnote{We generate $a_{\ell m}$`s with the \texttt{map2alm} \texttt{healpy} routine from the $Q, \ U$ maps. We convolve the maps with the new Gaussian beam and pixel window function following the method described in Appendix \ref{sec:a01}. Finally, we regenerate the $Q, \ U$ maps with the \texttt{alm2map} routine from the convolved $a_{\ell m}$`s.}. The smoothing process helps to increase the signal-to-noise ratio of the maps and to minimise any effect due to beam-asymmetries in the two experiments.

We estimate the noise level of the data at the $1^\circ$ and $3^\circ$ resolutions, including uncertainties due to smoothing and pixel downgrading. For \textit{WMAP}, we generate 600 Gaussian noise realisations based on the covariance matrices at full resolution. For \textit{Planck}, we use the 600 noise simulations provided on the PLA \cite{Npipe}. We downgrade and smooth each simulation in the same way as the data. Finally, for each pixel we compute $\sigma_Q^2$, $\sigma_U^2$ and $\sigma_{QU}^2$ from the variance and covariance over all of the simulated $Q$ and $U$ maps.

\subsection{Debiased Estimator}
\label{sec:02,sub:03}
\begin{figure}[htbp]
\centerline{\includegraphics[scale=.6]{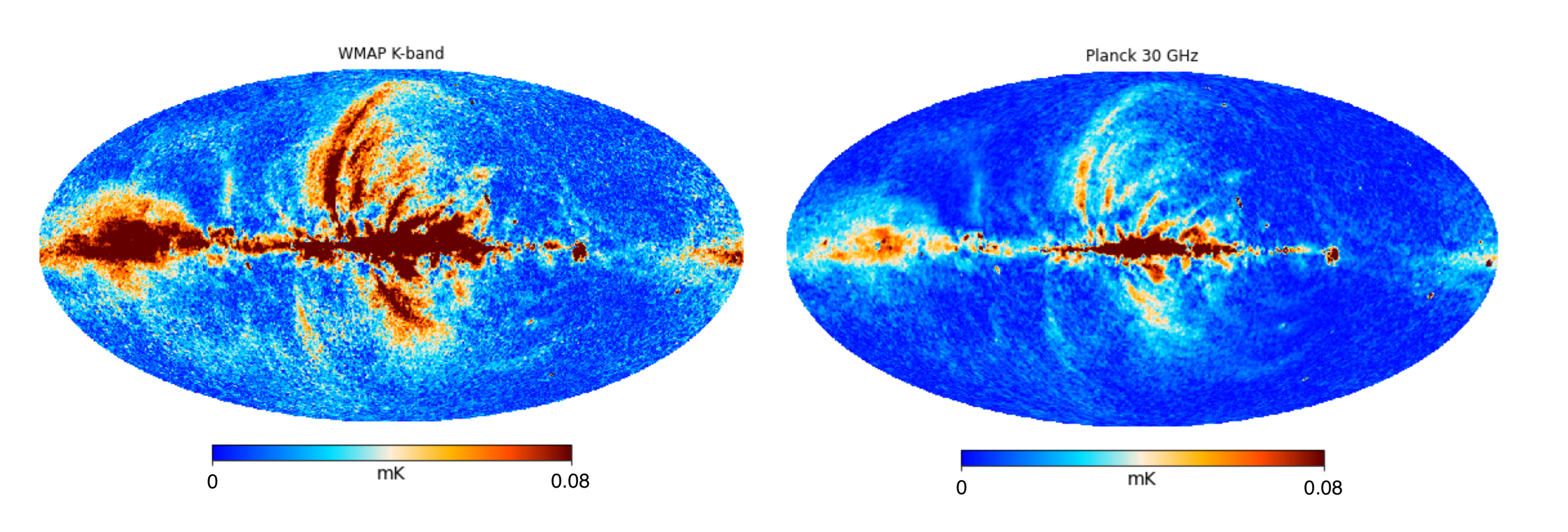}}
\caption{Debiased Polarized Intensity maps ($1^\circ$) of the \textit{WMAP} K-band at 23~GHz (left) and the 30~GHz channel of the \textit{Planck} PR4 data set (right). }
\label{fig:maps}
\end{figure}

A morphological analysis is applied to the polarized intensity $P = \sqrt{(Q^2+U^2)}$, which, given its positive nature, is subject to noise bias. In particular, in the low signal-to-noise regime, $P$ will yield a positive estimate even if $Q$ and $U$ are zero. 
We use the modified asymptotic (MAS) estimator \cite{Plaszczynski_2014} in order to correct the polarized amplitude for the bias. We recall that the debiased polarized amplitude with the MAS estimator is computed as
\begin{equation}
    P_{MAS} = P - \frac{1-\exp(-P^2/b^2)}{2P}b^2
\label{PMAS}
\end{equation}
where the noise bias $b$ is function of the pixel variance and the polarization angle $\phi = \arctan(U/Q)$ given by
\begin{equation}
    b^2 = \sigma^2_U \cos^2(\phi - \theta) + \sigma^2_Q \sin^2(\phi - \theta), \quad  \theta = \frac{1}{2}\arctan\left( \frac{\sigma_{QU}}{\sigma_Q^2-\sigma_U^2} \right).
\label{MESb}
\end{equation}
The maps are showed in figure~\ref{fig:maps}. \cite{Plaszczynski_2014} demonstrate that in the regime where the signal-to-noise ratio exceeds 2, the estimator is unbiased and essentially Gaussian. An estimate of the variance is then given by
\begin{equation}
    \sigma_P^2 = \sigma^2_Q \cos^2(\phi - \theta) + \sigma^2_U \sin^2(\phi - \theta).
\label{sigmaP}
\end{equation}
Note that the debiased polarized intensity and its variance are defined pixel by pixel. Thus, we do not take into account the correlation between pixels, which is introduced smoothing and downgrading of the maps.
Although we use the MAS estimator in our analysis, several tests have been performed using the Wardle $\&$ Kronberg estimator \cite{Wardle_1974, Vidal_2015, Vidal_2016} instead, finding consistent results. 

\section{Filament Finder}
\label{sec:03}

\subsection{Algorithm}
\label{sec:03,sub:01}
We develop an algorithm in order to seek elongated structures in polarized intensity maps. The method resembles the two-dimensional version of the Smoothed Hessian Major Axis Filament Finder (\texttt{SHMAFF}) \cite{bond2010acrawling, bond2010bcrawling}. The \texttt{SHMAFF} algorithm has previously been used to find filaments in the three-dimensional galaxy distribution and in the analysis of polarized dust structures in the \textit{Planck} 353~GHz maps \cite{PlanckFilaments}. 
The main difference in our implementation is that, while the \texttt{SHMAFF} detection is based on the minimal eigenvalues of the Hessian matrix, our detection is based directly on the polarized intensity. This is because the Hessian matrix fails to find pixels with minimal eigenvalues in detecting the thick and diffuse filaments which we expect to be in the noisier area of the sky, that is outside the Galactic plane. 

The algorithm works on a pixel by pixel basis, examining the orientation angle $\psi$ defined in \cite{PlanckFilaments} and comparing the polarized intensity value $P$ with respect to a threshold $P_{th}$. The orientation angle is determined from 
\begin{equation}
\begin{split}
    \psi &= \arctan\left( - \frac{H_{\theta\theta} - H_{\phi\phi} + \alpha}{2H_{\theta\phi}} \right) \\
   \textrm{with} &\ \alpha = \sqrt{(H_{\theta\theta} - H_{\phi\phi})^2 + 4H_{\theta\phi}^2},
\end{split}
\end{equation}
where $H$ is the Hessian matrix computed from the second-order covariant-derivatives with respect to the spherical coordinates $(\theta, \phi)$ \cite{Monteser_2005}. We compute the threshold values $P_{th}$ from the $P$ distribution. We cannot use the mean and the standard deviation because the distribution of $P$ is not Gaussian and exhibits an extended tail, thus, we define the threshold from the median $m_P$ and the median absolute deviation (MAD) $\sigma_m$ \cite{hampel1974influence, komm1999multitaper} as
\begin{equation}
    P_{th} = m_P + \sigma_m =  m_P + 1.4826 \cdot median(|P - m_P|).
\label{Pth}
\end{equation}

The algorithm starts by identifying the brightest pixel $P_{0}$ and denoting its orientation angle $\psi_{0}$. It then considers its 8 (or 7) neighbouring pixels, identified with the \texttt{get\_all\_neighbours} routine of \texttt{HEALPix}. For each neighbour pixel, two conditions are checked: $(i)$ if its polarized intensity is larger than the fixed threshold of equation \ref{Pth}, $(ii)$ if its orientation angle is coherent with the initial pixel
\begin{equation}
\begin{split}
(i) \ & \ P_i > P_{th} \\
(ii) \ & \ |\psi_i - \psi_0| < \Delta\psi,
\end{split}
\label{filMeth}
\end{equation}
where we fix $\Delta\psi$ at $10^\circ$ in order to detect the longest filaments. Nevertheless, we have checked that the results are robust for a reasonable range of $\Delta\psi$. 
If both conditions in equation~\ref{filMeth} are satisfied, we accept the pixel as part of the same structure of the initial pixel. Then, we look for its neighbours which also satisfy the two conditions. We continue this friend-of-friend recursive algorithm until the conditions are no longer satisfied. Once a structure is defined, the pixels are masked from the map, the new brightest pixel is identified and the condition-based procedure is repeated. 

We finally define a template, which includes all the pixels satisfying the previous conditions, with all the strongly polarized areas in the $P$ map of arbitrary shapes and sizes. We then smooth it with a $3^\circ$ Gaussian beam, in order to soften the boundaries. 
In order to allow only elongated structures which can be identified as filaments, we apply one further criteria to the $3^\circ$ smoothed map, i.e. we reject structures with length smaller than a threshold minimal length $L<L_{th}$, fixed at $10^\circ$. The length is defined as the maximum pixel-pair angular distance. The threshold length has been selected analysing simulations as discussed in appendix~\ref{sec:a02,sub:02}.

\subsection{Band-pass Filter}
\label{sec:03,sub:02}
Different filaments have been observed in low frequency (< 1~GHz) radio continuum surveys, and more recently in the \textit{WMAP} polarization data \cite{Vidal_2015}. The filaments can be divided into two categories:  bright and narrow, or weak and diffuse. 
In order to optimize the detection of these filaments, we filter the maps to focus on specific angular scales of interest before applying the filament finder algorithm. We compute the spherical harmonic coefficients $a_{lm}$ by means of the \texttt{map2alm} routine of \texttt{HEALPix}, multiply by a band-pass filter of the following form
\begin{equation}
    f(\ell) = \frac{1}{4}\left[ 1 + \tanh\left( \frac{\ell - \ell_{min}}{\Delta\ell} \right) \right]\left[ 1 - \tanh\left( \frac{\ell - \ell_{max}}{\Delta\ell} \right) \right]
\label{filter}
\end{equation}
then generate the filtered maps with the \texttt{alm2map} routine. The filter cuts off the amplitudes below a multipole scale $\ell_{min}$ and above $\ell_{max}$. In order to detect the thinner filaments, we consider multipoles in the range $20$-$50$, whereas for the diffuse filaments the multipoles are restricted to the range $15$-$20$. The cuts are roughly in accordance with the widths of the filaments ($\ell \sim 180/\theta)$. $\Delta\ell$ is set to 10, but the method is robust for a reasonable range of $\Delta\ell$. 

The filters are shown in figure~\ref{figfilter} in appendix~\ref{sec:02}. By filtering out small-scale modes, we enhance the contrast of larger structures with respect to the diffuse foreground emission, and also reducing the instrumental noise.
Moreover, we remove correlations on large scales which can negatively affect the detection. The application of the filter is critical to increase the accuracy of the estimation of the polarization orientations of the filaments, especially in areas where the signal-to-noise is low. 
A similar filter has been used in \cite{Rahman21} for a statistical analysis of the 408~MHz Haslam data.
    
\subsection{Results}
\label{sec:03,sub:03}
Starting from the debiased polarized intensity maps obtained as described in section~\ref{sec:02,sub:03}, we generate two bandpass-filtered maps $P^{20\text{-}50}$ and $P^{15\text{-}20}$, applying the filter in equation~\ref{filter}, where the superscripts correspond to the applied multipole ranges. 
We first mask the bright point sources, both Galactic and extragalactic, which could bias our algorithm. We use the mask derived for the \textit{Planck} PR4 SEVEM component separation pipeline which includes all the point sources that have polarization detection significance levels of 99\% or more in the 30~GHz polarized map \cite{Npipe, Argueso2009, Lopez2006}. 
The finder algorithm is based on the threshold condition in equation \ref{filMeth}, where the threshold is computed from the $P$ distribution (see equation \ref{Pth}). However, very bright areas such as the Galactic plane could bias the threshold value upwards, preventing filament detection. Therefore, when analysing the $P^{20\text{-}50}$ maps, we apply a Galactic mask excluding pixels at latitudes $|b|<3^\circ$. A similar argument applies to the $P^{15\text{-}20}$ analysis. However, since we are looking for very faint filaments, the NPS and the Southern Fan regions are also excluded in addition to the Galactic plane. This can be achieved by simply masking the brightest 30\% sky fraction. The specific choice of masks was tested on the simulations described in the appendix \ref{sec:a02,sub:01}.

\begin{figure}[htbp]
\centerline{\includegraphics[scale=.55]{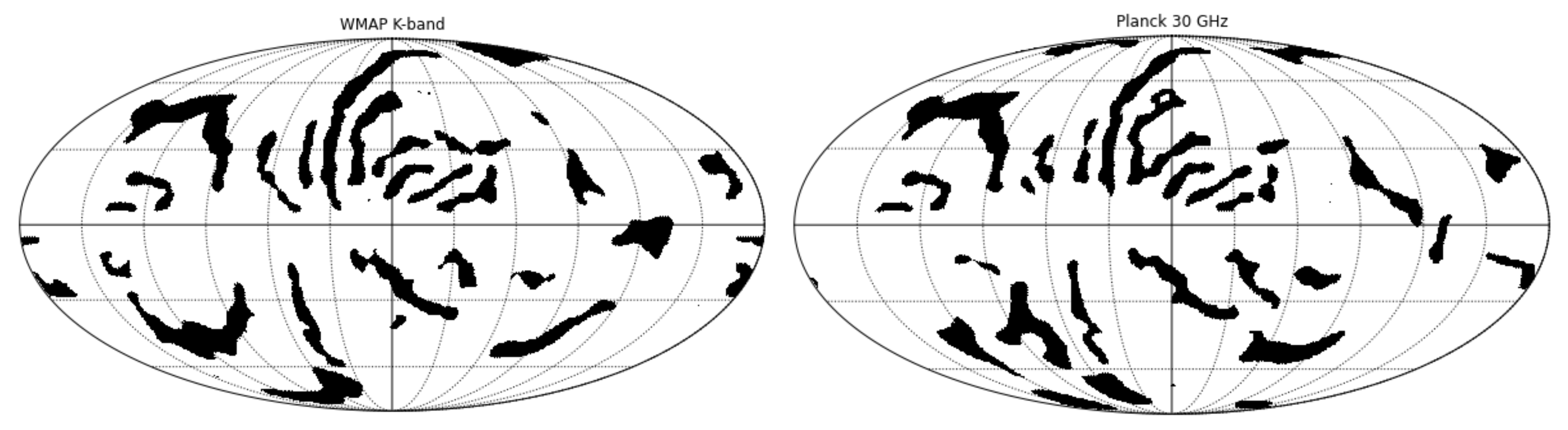}}
\caption{Filamentary structures detected in the debiased polarized intensity of the \textit{WMAP} K-band at 23~GHz (left) and the \textit{Planck} 30~GHz channel (right).}
\label{fig:filWP}
\end{figure}

After applying the filament finder algorithm to the two bandpass-filtered maps, we merge the two sets of results into one template, smooth it by $3^\circ$ and apply the criteria of minimal length as described in section~\ref{sec:03,sub:01}. The filament templates determined independently from the  \textit{WMAP} K-band and \textit{Planck} 30~GHz data are shown in figure~\ref{fig:filWP}. 
The \textit{WMAP} results reveal more compact and elongated structures than for \textit{Planck}, indicating that the finder algorithm performs better when applied to the data with brighter synchrotron emission.
However, several similar structures are detected in the same areas of the sky in both maps. The agreement between the independent results corroborates the validity of the algorithm and supports the existence of the filaments as real emission, and not due to noise or systematic effects. 
In appendix~\ref{sec:a02,sub:01}, we explore the accuracy and limits of the filament finder algorithm,
testing our method with toy filamentary foreground models. 

\section{Filament analysis}
\label{sec:04}

\subsection{Filaments}
\label{sec:04:sub:01}
\begin{figure}[htbp]
\centerline{\includegraphics[scale=.5]{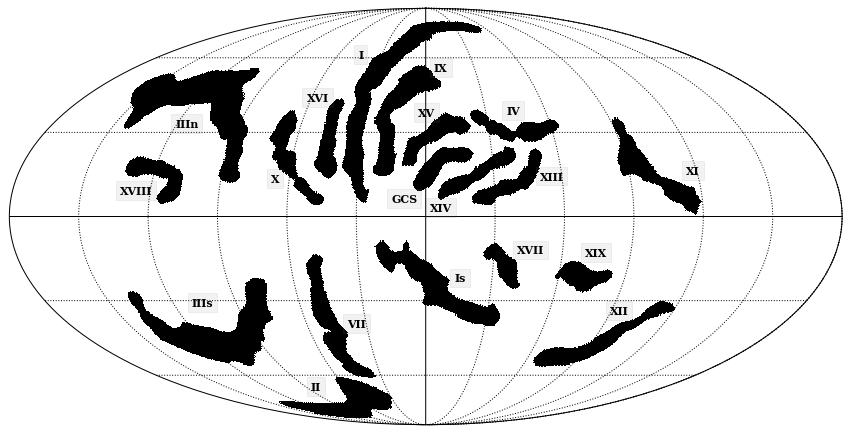}}
\caption{Template map showing the filaments of the polarized synchrotron emission detected by combining \textit{WMAP} (23~GHz) and \textit{Planck} (30~GHz) results.}
\label{fig:filaments}
\end{figure}

Combining the information obtained from \textit{WMAP} and \textit{Planck}, we get a final template of polarized filamentary structures detected in the frequency range 23-30 GHz, shown in figure~\ref{fig:filaments}. We use the \textit{WMAP} detection as our benchmark, but only retain those structures which are detected, at least in part, in the \textit{Planck} data. The one exception is filament XI, which is clearly detected in the \textit{Planck} data, but only partially in \textit{WMAP}. Its existence is supported by previous analysis performed on \textit{WMAP} \cite{Vidal_2015}. This method ensures that detected filaments are not due to noise or systematic residuals. Then, we remove those structures which do not show clear elongation, in particular in the Galactic plane, where the emission is more complex, and polar regions, where the signal-to-noise is low. 

In order to specify the filaments, we use and expand the nomenclature used in \cite{Vidal_2015}.
Filaments I (NPS), IIIn, and IV have been recognised and studied for more than 60 years. These large structures have been observed in X-ray, gamma-ray and other microwave experiments \cite{Brown60, Quigley65, Large66}. Filament II (Cetus Arc) was previously detected in the radio sky \cite{Large62}, and found here for the first time in polarization despite its low emission. We detect ten further filaments reported in \cite{Vidal_2015} (Is, GCS, IIIs, VII, IX, X, XI, XII, XIII, XIV) but not filament VIII. We also identify five new filaments (XV, XVI, XVII, XVIII, XIX) that are visible in both \textit{WMAP} and \textit{Planck}. Filament XV is a bright structure at the center of the Galactic Haze \cite{DickinsonSyn}. Filament XVI, because of the position and shape, seems to be related to Filament I. Filament XVIII is a bright structure of the Northern Fan region close to the Galactic plane. Finally, Filaments XVII and  XIX are new detections in the region below the Galactic plane. 

Most of the detected filaments have circular arc-like shapes, supporting the model of supernova remnants expanding into the Galactic magnetic field \cite{spoelstra1973galactic}. Several structures appear to be spatially correlated with each other, although most are stand-alone features. In the NPS, there are several elongated structures which do not resemble loop-like features. They were first identified in radio observations \cite{Large66}, but the NPS complexity is more evident in polarization. Because of their location, there are models which link these structures with the Fermi Bubbles (FB) detected in Fermi data at energies $\sim$10–500~GeV \cite{dobler2010fermi,Su_2010}. 

\subsection{Polarization Fractions}
\label{sec:04:sub:02}
The 
polarization fraction is defined as the ratio of the polarization amplitude to total intensity
\begin{equation}
    \Pi = \frac{P}{I}.
\end{equation}
At the low frequencies of interest here, the synchrotron emission largely dominates the polarization signal, thus, we can neglect other physical emission mechanisms and consider directly the frequency data. However, this assumption does not apply in intensity where the CMB, free-free and AME also contribute to the total emission. 

Most of the observations suggest that at frequencies above 20~GHz the spectral index of the synchrotron intensity spectrum is $\beta \approx -3$ \cite{PlanckXXV2016}. However, according to some models \cite{Orlando_2013}, it can get much flatter for frequencies below 10~GHz. Moreover, we expect it to exhibit significant spatial variations. 
In this analysis, we use the diffuse synchrotron intensity map provided in the Planck 2015 \cite{Planck2016} release. The template has been generated at a reference frequency of 408~MHz by \texttt{Commander} (a parametric component separation method) applied to the \textit{WMAP}, \textit{Planck} and \textit{Haslam} observations.  

We estimate the synchrotron intensity extrapolating the 408~MHz map up to 23~GHz and 30~GHz adopting a fixed spectral index $\beta = -3.0$. The maps are analysed at $3^\circ$ resolution and $N_{side}=64$, masking the Galactic plane\footnote{In order to mask the bright pixels along the Galactic plane, we use a Galactic mask obtained combining the 2015 Galactic plane mask which allows the 90 per cent of the sky (provided in the PLA) and a Galactic latitude mask excluding pixels within $\pm 5^\circ$ of the Galactic plane.}. Pixels with a signal-to-noise ratio lower than 2.5 have been excluded. 

The polarization fraction maps are showed in figure~\ref{figpolFracBeta} (top panels). For each pixel, the polarization fraction error, $\sigma_\Pi$, is obtained propagating the errors in the polarization and intensity maps. The intensity uncertainty is, in turn, obtained propagating the uncertainty of the intensity spectral index. The weighted average of the polarization fractions over pixels for each filament are listed in Table~\ref{table1}. The largest source of uncertainty in $\sigma_\Pi$ is due to the uncertainty in the intensity spectral index. We are aware that, locally, $\beta$ can assume values over a very broad range. However, we compute the polarization fraction averaging over extended areas, so it is reasonable to assume that in these areas $\sigma_\beta = 0.1$ as found in previous works on partial-sky analysis \cite{planck2016PS,Martire_2022,QJ_Elena}. 
We report good agreement between the \textit{WMAP} and \textit{Planck} results, the largest discrepancies arising for the more diffuse filaments, e.g., filament XII.

Assuming a uniform spectral index, we find that the polarization fraction of the filaments are typically larger than for external regions outside the filaments. The filaments with the highest polarization fractions, IX and XV, achieve values above 30\%, and are both located in the NPS. Loop I has an average value of about 20\%, slightly smaller than the value found at the center of the NPS. These results corroborate the results found in the previous analysis \cite{Vidal_2015, PlanckXXV2016}. We report a high polarization fraction also for filament XVIII, located in the Fan region. The lowest polarization fractions are found for filaments XI, XIII and XVII.

\subsection{Spectral Index}
\label{sec:04:sub:03}
The synchrotron spectral energy distribution (SED) is generally approximated by a power law\footnote{Given the sensitivity of \textit{Planck} and \textit{WMAP} data, we cannot explore more complex models.} $S_\nu \propto \nu^\beta$ where $\beta$ is the energy spectral index. Spatial variations of $\beta$ have been reported in the literature \cite{Svalheim2020BeyondPlanckXP,2022MNRAS.517.2855D, QJ_Elena}. In this section, we measure the spectral index of each filament described in \ref{sec:04:sub:01} employing a method based on the $Q$ and $U$ Stokes parameters \cite{Fuskeland2014,FuskelandSynch}. 

Let us define the vector
\begin{equation}
    d(\alpha) = Q \cos(2\alpha) + U \sin(2\alpha)
\label{eq:d}
\end{equation}
which represents, for each pixel, the projection of the Stokes parameters $(Q,U)$ into a reference frame rotated by the angle $\alpha$. We vary $\alpha$ over the range $(0^\circ, 85^\circ)$, in steps of $5^\circ$. For each filament, we compute a linear fit over all the internal pixels to the relation
\begin{equation}
d_{_{\textrm{P30}}}(\alpha) = m(\alpha) \cdot d_{_{\textrm{WK}}}(\alpha) + n(\alpha)
\label{eq:SEDline}
\end{equation}
where $d_{_{\textrm{P30}}}(\alpha)$ and $d_{_{\textrm{WK}}}(\alpha)$ are computed respectively from the \textit{Planck} and \textit{WMAP} data. Adding the free parameter $n(\alpha)$ gives the advantage of removing any zero level due to possible systematics in the maps. The fit is performed with the orthogonal distance regression code \texttt{odr}\footnote{https://docs.scipy.org/doc/scipy/reference/odr} from \texttt{SciPy}, in order to account for the noise variance of both \textit{WMAP} and \textit{Planck}. A calibration error of 0.3\%  has been added in quadrature to both experiments \cite{Bennett_2013,Npipe}. 
From the parameter $m(\alpha)$ and its uncertainty, we compute the spectral index for each $\alpha$ as
\begin{equation}
\beta(\alpha) =\frac{\log m(\alpha)}{\log(\nu_{_{\textrm{P30}}}/\nu_{_{\textrm{WK}}})} \quad  \sigma_\beta(\alpha) =\frac{\sigma_m(\alpha)}{m(\alpha)}\frac{1}{\log(\nu_{_{\textrm{P30}}}/\nu_{_{\textrm{WK}}})}
\label{eq:SEDbetaAlpha}
\end{equation}
where $\nu_{_{\textrm{P30}}}=$28.4~GHz and $\nu_{_{\textrm{WK}}}$=22.8~GHz. The final value of the index is recovered from the weighted average 
\begin{equation}
\beta = \frac{ \sum_{\alpha=0}^{85} \beta(\alpha) \sigma_\beta^{-2}(\alpha)}{\sum_{\alpha=0}^{85} \sigma_\beta^{-2}(\alpha)}.
\label{eq:SEDbeta}
\end{equation}
Since the $\beta(\alpha)$ values are strongly correlated, we take as the uncertainty on the final spectral index the minimum variance among the measurements $\sigma_\beta = \min(\sigma_\beta(\alpha))$. We checked that this uncertainty is always larger then the intrinsic uncertainty of $\beta(\alpha)$ given by the standard deviation estimated at different rotation angles. 

As a consistency check, we also fitted the spectral index from the debiased polarized intensities with the T-T plot approach. We find consistent results with those determined with the method presented above. It has been shown that the synchrotron spectral index is not stable with respect to polarization orientation in the presence of systematics \cite{2013ApJ...763..138W}. The $(Q,U)$ method allows marginalization of the result over the polarization angle, making this approach more reliable. 

The maps are analysed at $3^\circ$ resolution and $N_{side}=64$, masking the Galactic plane. Pixels with a signal-to-noise ratio lower than 2.5 have been excluded. The maps are converted to Rayleigh-Jeans temperature units and corrected by the colour correction using the coefficients given in \cite{planckDust}. 

\begin{figure}[htbp]
\centerline{\includegraphics[scale=.6]{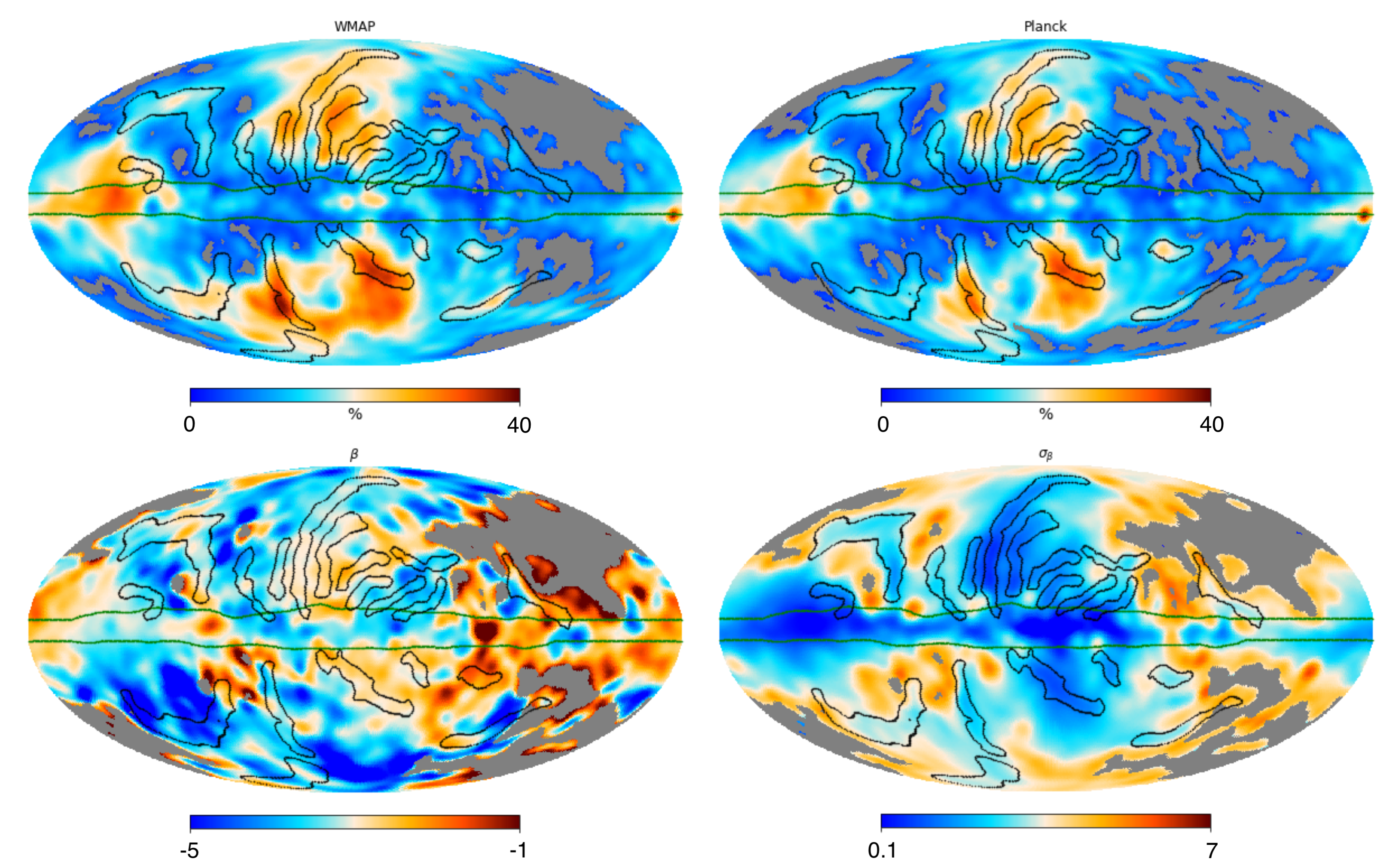}}
\caption{Top: polarization fraction of the \textit{WMAP} K-band at 23~GHz (left) and the 30~GHz channel of the \textit{Planck} PR4 (right). Bottom: spectral index $\beta$ (left) with error $\sigma_\beta$ (right). Outlines of the filaments (black) and the Galactic mask (green) used in the analysis are also shown. Pixels with a signal-to-noise ratio lower than 2.5 have been masked (grey). For illustrative purpose, the maps are smoothed to a resolution of $5^\circ$.}
\label{figpolFracBeta}
\end{figure}

In Table~\ref{table1} the spectral indices determined for the different filaments are listed. For each filament, a mean $\chi^2$ value is computed by averaging over all the values given at different rotation angles by the \texttt{odr} routine. Note that we do not take into account the presence of correlated noise between pixels, thus, leading to an underestimation of the uncertainties. This is one of the reasons which would explain some large $\chi^2$ values.
We find spectral index results, both inside and outside the filaments, consistent with the values found in previous analysis of about –3.1 \cite{akrami2020planck,NicolettaSych, Martire_2022}. As shown in figure~\ref{figpolFracBeta} (bottom panel), the $\beta$ values span a very broad range, from -3.59 (XVII) to -2.17 (VII). The value for loop I is consistent with the literature \cite{Vidal_2015, Guidi}. Filaments IX and XV, the most polarized detections located at the NPS, show slightly flatter values ($\sim$ -2.5). 

\begin{table}[htp]
\centering
\begin{tabular}{c|c||P{1cm}|P{1.2cm}|P{1cm}|P{1.2cm}||P{1cm}|P{1cm}|P{1cm}}
 \multicolumn{2}{c||}{  }  & \multicolumn{4}{c||}{polarization fraction} & \multicolumn{3}{c}{spectral index} \\
\hline
 \multicolumn{2}{c||}{  }  & \multicolumn{2}{c|}{\textit{WMAP}} & \multicolumn{2}{c||}{\textit{Planck}} & \multicolumn{3}{c}{} \\
\hline
 \rule{0pt}{3ex}
   filament & $f_{sky}$ [\%] & $\Pi$ [\%] & $\sigma_\Pi$ [\%] & $\Pi$ [\%] & $\sigma_\Pi$ [\%] & $\beta$ & $\sigma_\beta$ & $\chi^2_r$ \\ [1ex] 
  \hline
   \rule{0pt}{3ex}
I       &    1.5 &     22.6 &       9.4 &     21.2 &       9.3 & -3.06 &   0.01 &  1.92 \\
 Is     &    1.2 &     18.3 &       7.6 &     19.4 &       8.5 & -2.80 &   0.03 &  1.87 \\
II      &    0.8 &     24.5 &      10.4 &     20.4 &       9.2 & -3.37 &   0.06 &  3.01 \\
IIIn    &    2.1 &     18.9 &       8.0 &     17.4 &       7.7 & -3.07 &   0.02 &  1.55 \\
IIIs    &    1.8 &     19.8 &       8.4 &     15.5 &       7.1 & -3.65 &   0.04 &  2.22 \\
IV      &    0.5 &     18.1 &       7.6 &     15.1 &       6.8 & -3.07 &   0.11 &  1.95 \\
GCS     &    0.4 &     22.6 &       9.4 &     22.0 &       9.6 & -2.80 &   0.03 &  2.35 \\
VII     &    0.8 &     17.3 &       7.3 &     20.4 &       9.1 & -2.17 &   0.09 &  1.97 \\
IX      &    0.9 &     33.1 &      13.7 &     32.2 &      14.1 & -2.64 &   0.02 &  1.90 \\
X       &    0.6 &     17.1 &       7.2 &     14.2 &       6.3 & -3.39 &   0.04 &  1.48 \\
XI      &    0.8 &     13.2 &       5.7 &     13.3 &       6.0 & -2.45 &   0.05 &  1.60 \\
XII     &    0.7 &     20.6 &       9.0 &     13.9 &       6.5 & -3.38 &   0.05 &  1.21 \\
XIII    &    0.5 &     14.0 &       5.8 &     12.6 &       5.6 & -3.28 &   0.03 &  2.42 \\
XIV     &    0.5 &     17.6 &       7.3 &     14.5 &       6.4 & -3.05 &   0.04 &  1.34 \\
XV      &    0.5 &     33.8 &      14.0 &     34.8 &      15.3 & -2.36 &   0.04 &  2.44 \\
XVI     &    0.4 &     22.0 &       9.1 &     20.7 &       9.1 & -3.39 &   0.11 &  1.51 \\
XVII    &    0.3 &     14.6 &       6.1 &     15.7 &       6.9 & -3.59 &   0.05 &  1.67 \\
XVIII   &    0.5 &     30.1 &      12.5 &     26.4 &      11.6 & -3.08 &   0.05 &  1.28 \\
XIX     &    0.4 &     21.1 &       9.1 &     26.1 &      11.6 & -2.28 &   0.09 &  0.91 \\
 \hline
 \hline
    \rule{0pt}{3ex}
inside filaments  &   15.2 &     19.0 &       8.0 &     17.2 &       7.7 & -3.08 &   0.01 &  3.14 \\
outside filaments &   62.1 &     11.1 &       5.0 &     10.3 &       4.9 & -3.15 &   0.01 &  3.34 \\
combined &   76.2 &     11.8 &       5.2 &     11.0 &       5.2 & -3.10 &   0.01 &  3.29 \\

\hline
\end{tabular}
\caption{Polarization fractions and polarization spectral indices of the \textit{WMAP} K-band and \textit{Planck} 30~GHz channel. The synchrotron intensity map is extrapolated from the 408~MHz map up to 23~GHz and 30~GHz using a constant spectral index $\beta = -3.0$. The spectral index is computed over the 23-30~GHz frequency range.}
\label{table1}
\end{table}

\subsection{$E$ and $B$ Modes}
\label{sec:04,sub:04}

As described in appendix~\ref{sec:a01}, the polarized emission can be decomposed into $E$ and $B$ modes. The synchrotron polarized angular power spectra has been analysed in the frequency range 2-30~GHz, finding that: both $E$- and $B$- modes can be well described by a power law $C_\ell^{EE,BB} \propto \ell^{-2.9}$, the $B$-to-$E$ ratio ranges between 0.2-0.5 and the $EB$ correlation is compatible with zero \cite{NicolettaSych,Martire_2022}. In this section we analyze how the filamentary structure relates to the two polarized components.

In \cite{Liu_2018} a method was proposed for decomposing the $Q$ and $U$ Stokes parameters into the so-called $E$- and $B$-mode families. Starting from the ($Q$, $U$) maps, we can compute the $a^{E,B}_{lm}$ coefficients using the \texttt{map2alm} routine of \texttt{HEALPix}. 
Setting $a^B_{lm}=0$ and computing the Stokes parameters with the \texttt{alm2map} routine determines the contribution to $Q$ and $U$ from the $E$ mode alone. Similarly, setting $a^E_{lm}=0$, we get the contribution from the $B$ mode. Therefore, we can compute the single-mode polarization intensities as
\begin{equation}
    P_E = \sqrt{Q_E^2 + U_E^2} \quad  P_B = \sqrt{Q_B^2 + U_B^2}.
\end{equation}

For the sake of brevity, here we show an analysis performed on the \textit{WMAP} K-band, however, the same conclusions can be obtained from the 30~GHz \textit{Planck} data. We do not use any estimators to correct the polarized amplitude for the noise bias. This is because the estimator in equation \ref{PMAS} would require the decomposition of the variance into the $E$- and $B$ families, which is not a straightforward operation. Nevertheless, we expect qualitatively correct results for the areas where the signal-to-noise is high, in which the bias is negligible.

\begin{figure}[htbp]
\centerline{\includegraphics[scale=.58]{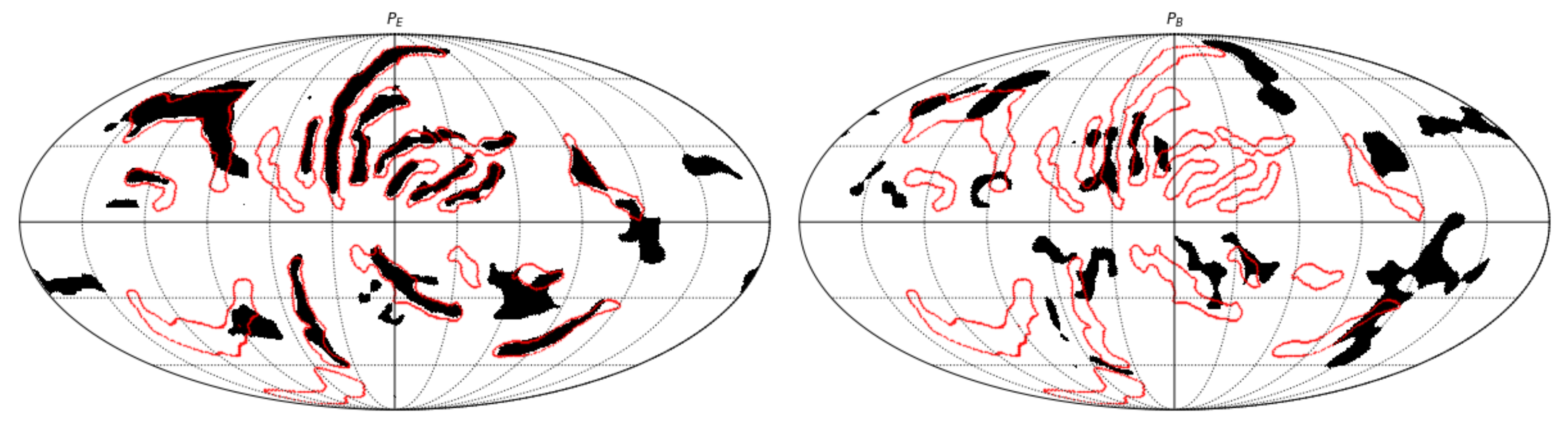}}
\caption{Filamentary structures detected in the decomposed polarized intensity $P_E$ (left) and $P_B$ (right) maps obtained with the \textit{WMAP} K-band at 23~GHz. In red, the filaments detected from the polarized intensity $P_{MAS}$ as described in section~\ref{sec:04}. }
\label{fig:filEB}
\end{figure}

Applying the filament finder algorithm to the decomposed maps, we find that nearly all of the filamentary structures are detected at least in part in the $P_E$, but not in the $P_B$ map, as shown in figure~\ref{fig:filEB}. As expected, the algorithm fails to detect filaments II, IIIs and X, which are either diffuse or strongly affected by noise bias. The complex filamentary structure of the NPS emission is clearly visible in $P_E$, suggesting its $E$-nature. In the $P_B$ maps, we detect parts of filaments I and IX (even if slightly shifted). Filament XIX is the only structure which is partially detected in $B$, but not in $E$. 

\subsection{408~MHz Haslam Map}
\label{sec:04,sub:05}
The current best full-sky map of the synchrotron intensity emission at 408~MHz is due to \cite{haslam1981408, haslam1982408}. A more recent version of the data that has been destriped and cleaned of bright point sources is described in \cite{remazeilles2015improved}. We apply the filament finder algorithm to this map, provided as a LAMBDA product\footnote{https://lambda.gsfc.nasa.gov/product/foreground/fg\_2014\_haslam\_408\_info.html}, to study the filamentary detection in intensity at low frequencies. 

\begin{figure}[htbp]
\centerline{\includegraphics[scale=.58]{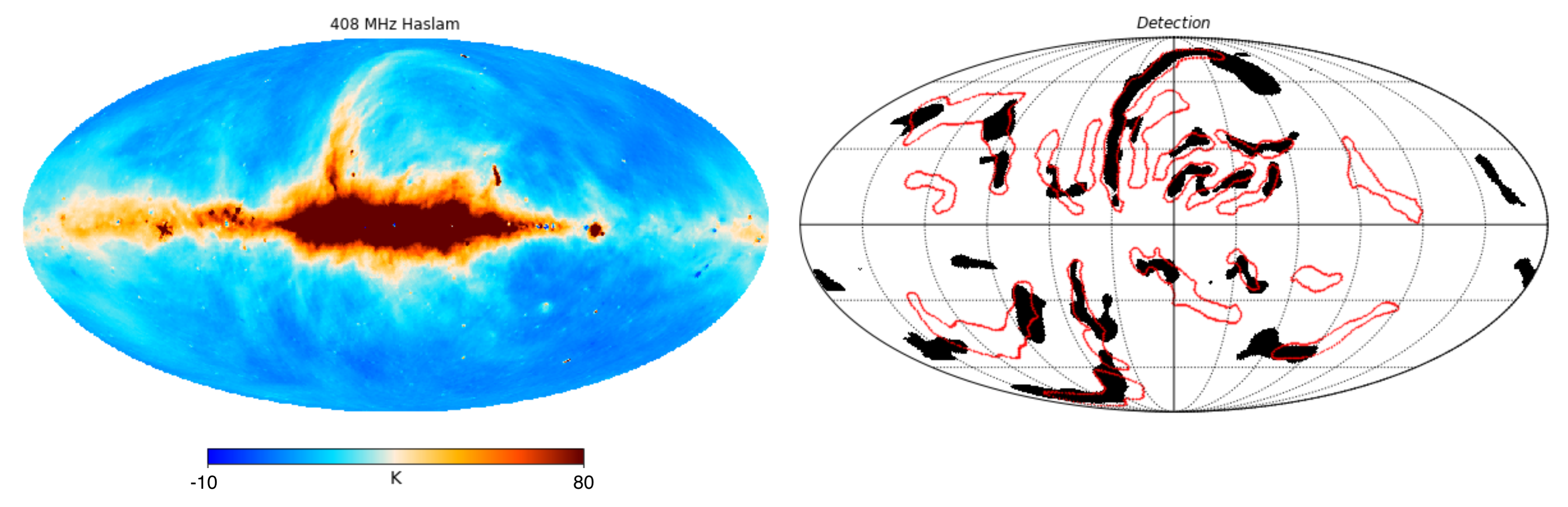}}
\caption{Left: full-sky 408~MHz map. Right: Structures detected in the data (black) compared to the filaments found in polarization at 23-30~GHz (red).}
\label{fig:filInt}
\end{figure}

The map and the detected filaments are shown in figure~\ref{fig:filInt}. Filament I is the major structure in intensity as well as in polarization. Part of the detected structure is well matched by what is found at 23-30~GHz, but is more extended at the lower frequency. A similar observation holds for filament II. The majority of the structures in the NPS observed with \textit{WMAP} are also visible, at least in part, in the \textit{Haslam} map. However, in intensity the synchrotron emission is very diffuse, and the algorithm fails to detect the more diffuse filaments, such as III, XI, XII. An interesting result is that filaments IX and XVI, which are the two strong emission structures detected in polarization around Loop I, are not well detected in the 408~MHz map. 

Note that the analysis presented in this section provides us with an additional robustness test for our algorithm. However, only a qualitative comparison between \textit{Haslam} and \textit{WMAP}/\textit{Planck} maps is possible. This is because, in intensity, the synchrotron emission dominates at 408~MHz, but not at 23/30~GHz, where other components become important. On the other hand, the synchrotron emission dominates in polarization, but comprehensive data sets in polarization do not exist at 408~MHz. We hope that the results presented in this section will stimulate future analysis using new data at similar frequencies.

\section{Statistical Properties}
\label{sec:05}

\subsection{Minkowski Formalism}
\label{sec:05,sub:01}
According to Hadwinger’s Theorem, any morphological property can be expressed as a linear combination of Minkowski Functionals (MFs). These are defined for any field not requiring any prior assumption, making them particularly advantageous for the analysis of fields for which a non-Gaussian nature is known. Several analyses have already been performed on CMB data using MFs to search for evidence of non-Gaussianity \cite{Schmalzing_1998, Chingangbam_2017, Carones22} and residual foreground contamination \cite{Chingangbam_2013}, or to characterize the properties of foregrounds \cite{Rana_2018,Rahman21}. In this section, we will briefly review the method used for their numerical calculation, following the methodology developed by \cite{Chingangbam1_2017}.

Given a map $u$ on the sphere ($\mathbb{S}^2$) and a threshold $v$, there are 3 MFs which represent the area ($V_0$), the perimeter ($V_1$) and the integrated geodesic curvature ($V_2$) of an excursion set, that is the region where $u\geq v$, with boundaries defined by $u=v$.
For a map in the \texttt{HEALPix} pixelization, we can numerically compute the MFs via a sum over all pixels
\begin{equation}
    V_0(v) = \frac{1}{N_{pix}} \sum_{pixels}\mathcal{H}(u-v)
\label{V0}
\end{equation}
\begin{equation}
    V_1(v) = \frac{1}{4}\frac{1}{N_{pix}} \sum_{pixels}\delta(u-v)\sqrt{u_{;\theta}^2 + u_{;\phi}^2}
\label{V1}
\end{equation}
\begin{equation}
    V_2(v) = \frac{1}{2\pi}\frac{1}{N_{pix}} \sum_{pixels}\delta(u-v) \frac{u_{;\theta}u_{;\phi}u_{;\theta\phi} - u_{;\theta}^2u_{;\phi\phi}  - u_{;\phi}^2u_{;\theta\theta}}{u_{;\theta}^2 + u_{;\phi}^2}
\label{V2}
\end{equation}
where $u_{;i}, u_{;ij}$ ($i, j \in (\theta, \phi)$) are the first and second partial derivatives in spherical harmonic space, $\mathcal{H}$ is the Heaviside step function and $\delta$ is the delta function. We rescale the field $u$ to have zero mean and unit standard deviation. 
The $\delta$-function is numerically approximated through a discretization
\begin{equation}
\delta(u-v) = \frac{1}{\Delta v} \left[\mathcal{H}\left(u+\frac{\Delta v}{2} \right) - \mathcal{H}\left(u-\frac{\Delta v}{2} \right)\right]
\label{delta}
\end{equation}
that is $\delta(u-v)$ is equal to $1/\Delta v$ if $u$ is between $v-\Delta v/2$ and $v+\Delta v/2$, and zero elsewhere. 

This pixelization method introduces systematic residuals. It has been shown that residuals scale as the square of the bin-size (${\Delta v}^2$) \cite{Lim_2012}. However, if the bin-size is too small the results can be affected by map noise. We pick the value $\Delta v = 0.5$. We find that this bin-size minimizes the residual obtained comparing the numerical equations \ref{V1} and \ref{V2} with the MFs analytical equations valid for a perfect Gaussian field. 

Minkowski Tensors (MTs) are tensorial quantities that generalize the scalar MFs. MTs have been already used in cosmology to study CMB \cite{Joby_2019} and foreground \cite{Rahman21} anisotropies. There are three rank-two MTs on the sphere, usually denoted as $W_k$. The three scalar MFs, $V_k$, are then given by the traces of $W_k$. We are particularly interested in $W_1$, also called the Contour Minkowski Tensor (CMT), which encodes shape and alignment information for structures. It can be numerically computed as
\begin{equation}
    W_1(v) = \frac{1}{4}\frac{1}{N_{pix}} \sum_{pixels} \delta(u-v) \frac{1}{|\nabla u|} M
\label{W1}
\end{equation}
where 
\begin{equation}
M = 
\begin{pmatrix}
u_{;\phi}^2 & - u_{;\phi}u_{;\theta} \\
- u_{;\phi}u_{;\theta} & u_{;\theta}^2
\end{pmatrix}.
\label{M}
\end{equation}
$W_1$ is proportional to the identity matrix if the structures have no elongation in any particular direction. $\lambda_+$, $\lambda_-$ are the two eigenvalues of $W_1$ such that $\lambda_+$ > $\lambda_-$. We define $\alpha$ as
\begin{equation}
    \alpha = \frac{\lambda_+}{\lambda_-}.
\label{alpha}
\end{equation}
$\alpha=1$ implies that the field preserves statistical isotropy (SI). 
In order to quantify the non-Gaussianity and anisotropy of data maps, we need to compare them with a set of suitable simulations. We then define, at each threshold $v$, the quantities
\begin{equation}
    \chi_k = \frac{|\Delta V_k|}{\sigma_{V_k}}, \quad  \Delta V_k(v) = V_k^{data}(v) - V_k^{sim}(v)
\label{chiMF}
\end{equation}
where $V_k^{data}$ is the $k$-th MF computed from the data, $V_k^{sim}$ and $\sigma_{V_k}$ are the values obtained taking respectively the average and the standard deviation of the functionals determined from simulations. Analogously, we define at each threshold $v$ the quantities
\begin{equation}
    \chi_{(W_1)_{ii}} = \frac{|\Delta (W_1)_{ii}|}{\sigma_{(W_1)_{ii}}}, \quad  \Delta (W_1)_{ii} =(W_1)_{ii}^{data} - (W_1)_{ii}^{sim}
\label{chiMT}
\end{equation}
where $(W_1)_{ii}$ stands for $(W_1)_{11}$ and $(W_1)_{22}$, which are the diagonal terms of the CMTs. The values $(W_1)_{ii}^{sim}$ and $\sigma_{(W_1)_{ii}}$ are computed from simulations. The same quantification could not be applied to $\alpha$ because its statistic follows the Beta probability distribution \cite{Rahman21}. Thus we define the quantity 
\begin{equation}
    \chi_{\alpha} = \frac{\Delta\alpha}{\delta_1}\mathcal{H}(-\Delta\alpha) + \frac{\Delta\alpha}{\delta_2}\mathcal{H}(\Delta\alpha), \quad  \Delta \alpha(v) = \alpha^{data}(v) - \alpha^{sim}(v)
\label{chialpha}
\end{equation}
where $\delta_1$ and $\delta_2$ denote the 95\% confidence interval and $\mathcal{H}$ is the Heaviside step function. Note that a value $|\chi_{\alpha}|>1$ implies a deviation from the simulations outside the 95\% confidence interval. 
In this and the following sections we will focus on the analysis and results obtained from the \textit{WMAP} K-band polarization maps at resolution $1^\circ$ and $N_{side}=128$. The same analysis has been performed with 30~GHz \textit{Planck} maps and the results presented in appendix~\ref{sec:a03}.

\subsection{Masking and Filtering}
\label{sec:05,sub:02}
It is well known that the synchrotron emission is non-Gaussian and anisotropic on the full sky. In this analysis, we are mainly interested in understanding if this behaviour holds in regions of the sky where the emission is more diffuse, that is when the brightest areas (Galactic plane, the Spur and Fan regions) are masked. 

We define two different masks for the analysis of the faintest 80\% and 60\% of the sky. To avoid possible leakage of power which can arise when computing power spectra in the presence of sharp boundaries between the masked and the unmasked regions, we apodize the masks with the \texttt{mask\_apodization} ("C2") routine of \texttt{NaMaster}\footnote{\texttt{NaMaster} is a public software package providing a general framework to estimate  pseudo-$C_\ell$ angular power spectra. \cite{namaster} namaster.readthedocs.io} with apodization scale of $5^\circ$. We also take into account a point source mask, as used in the filament finder analysis in section~\ref{sec:03,sub:03}, apodized at $1^\circ$. To minimize the effects of the mask boundary in the sum in equations~\ref{V0}-\ref{W1}, we only include those pixels such that the smoothed mask value is larger than 0.9. For the two masks adopted here, this corresponds to sky fractions of 76.4\% and 57.1\%. The threshold has been conservatively chosen in order to maximize the statistical significance of our results. Note that we do not construct a specific filament mask, because from prior tests we have noticed that the Minkowski method is more reliable when we use a compact mask rather than a complex mask with many holes and islands.

\begin{figure}[htbp]
\centerline{\includegraphics[scale=.55]{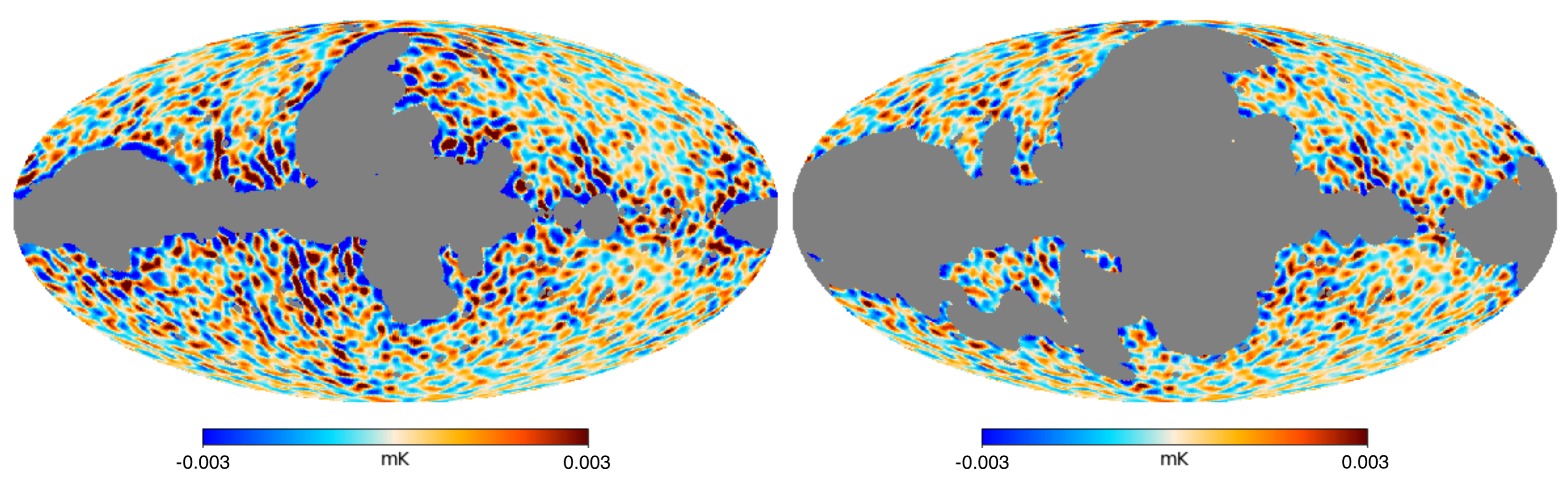}}
\caption{ Debiased \textit{WMAP} K-band polarized intensity maps at $1^\circ$ resolution, after band-pass filtering as defined in equation~\ref{filter} with $\ell_{min}=30$ and $\ell_{max}=180$. The grey regions correspond to the missing pixels for the 80\% (left) and 60\% (right) masks.}
\label{fig:maskFilMaps}
\end{figure}

Besides the region-dependency, in this analysis we want to test the statistical behavior on different scales. The maps are therefore filtered using the band-pass filter presented in section~\ref{sec:03,sub:02}. We analyze different scale ranges varying $\ell_{min}$ in equation~\ref{filter}. We do not study the maps for multipoles smaller than $\ell_{min}=30$ because the anisotropic nature of the emission at such large scale is well known, so we will only focus on $\ell_{min}\geq30$. We fix $\ell_{max}=180$, in accordance with the map resolution at $1^\circ$. In figure~\ref{fig:maskFilMaps} the \textit{WMAP} polarized intensity maps, filtered with the ($\ell_{min}=30$, $\ell_{max}=180$) band-pass, are shown. The region in grey corresponds to the pixels masked in the Minkowski analysis. 
We finally subtract the mean and then rescale with the standard deviation, where mean and standard deviation are computed from the data in the unmasked area of the maps. 

\subsection{Gaussian Simulations}
\label{sec:05,sub:03}
In order to quantify the non-Gaussianity and anisotropy of the polarized synchrotron emission, we need to compare the polarization data with a set of suitable simulations. To generate simulations, we first compute with \texttt{NaMaster} the polarization power spectra of the data maps in the unmasked areas. In fact, we compute the cross-spectra between the co-added \textit{WMAP} 1 to 4 year sky maps and 5 to 9 year data. In this way, we reduce the effects of instrumental noise and systematics. 
Based on this spectrum, we then generate Gaussian and isotropic simulations of full-sky $Q$ and $U$ Stokes parameter maps at the data resolution using the \texttt{HEALPix} \texttt{synfast} routine. We add noise and compute the debiased polarized intensity of the simulations as described in section~\ref{sec:02,sub:03}. 
We obtain 600 total polarization simulations which have a mean power spectrum compatible with the data, but are isotropic and generated from Gaussian-distributed $Q$ and $U$. Note that the variance of the simulations includes both sample variance and noise. Finally, we compute the MFs and MTs for the simulations using the same band-pass filter and masks as used for the data. 

\subsection{Results}
\label{sec:05,sub:04}
\begin{figure}[htbp]
\centerline{\includegraphics[scale=.73]{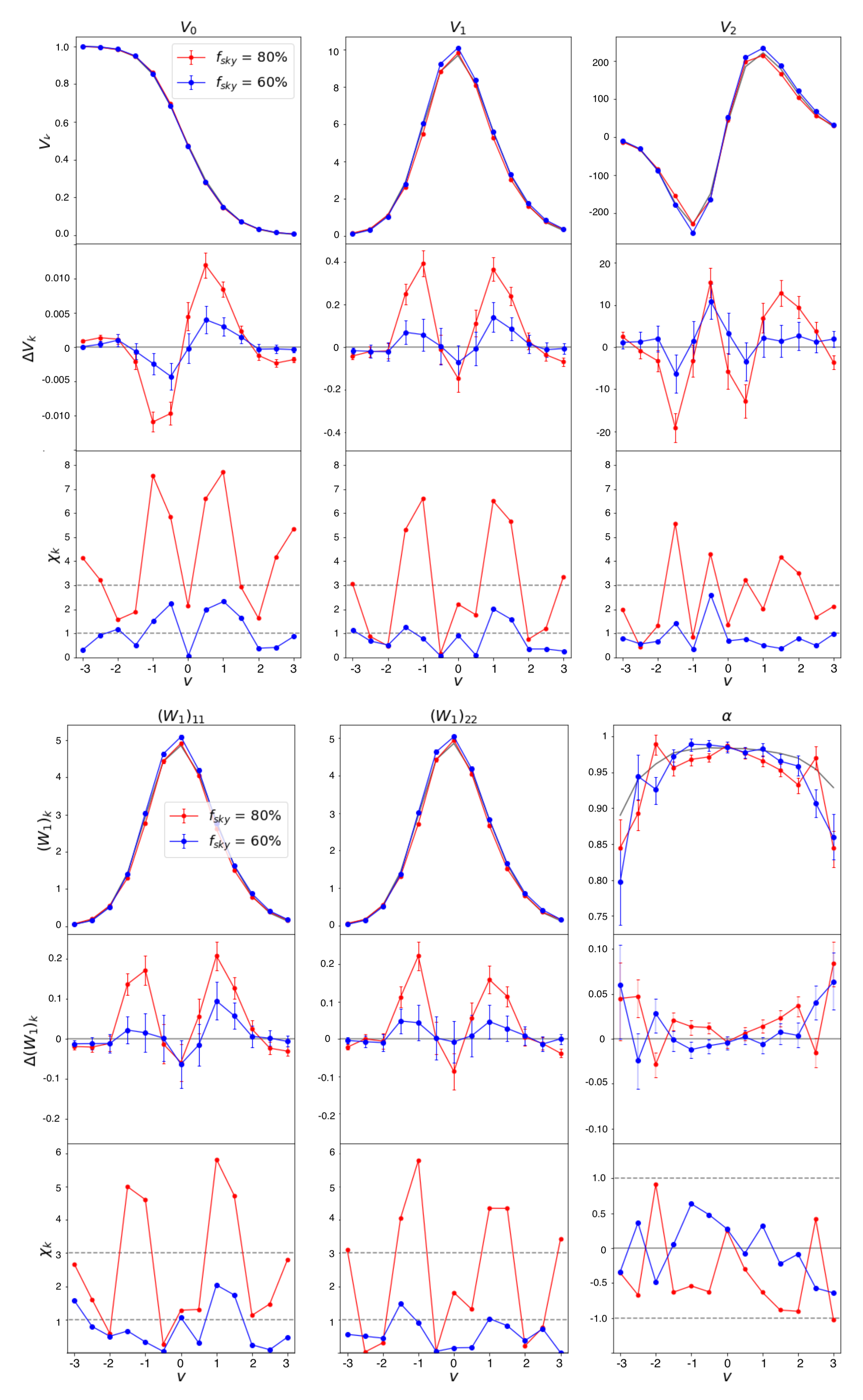}}
\caption{Upper panel: MFs (top), difference between data and Gaussian simulations (middle) and $\chi_k$ (bottom) as a function of threshold. Lower panel:  CMT diagonal terms and $\alpha$ (top), difference between data and Gaussian simulations (middle) and $\chi_k$ (bottom) as a function of threshold. The maps are previously filtered with a band-pass ($\ell = 30-180$), error bars are computed from simulations.}
\label{fig:MT30}
\end{figure}

The results for the MFs and the CMT, derived from band-pass filtered data are shown in figure~\ref{fig:MT30}. We find that the non-Gaussian and anisotropic deviations are much lower when bright regions are masked. For the larger sky fraction ($f_{sky}=80\%$), we find that the first MF ($V_0$) shows the largest deviation from Gaussianity. For most thresholds, the deviation is higher than $3\sigma$ and several values exceed $5\sigma$. The deviations of the other two MFs and the CMT diagonal terms exceed $3\sigma$ for several thresholds, few exceed $5\sigma$. $|\chi_{\alpha}|$ is lower than one except for one threshold, implying a weak deviation from isotropy. For the smaller sky fraction ($f_{sky}=60\%$), for all the thresholds we find that the deviations of the three MFs and the CMT diagonal terms never exceed $3\sigma$, with average values (over all thresholds) lower than $1.2\sigma$. The value of $\alpha$ never exceeds the 95\% interval. 

\begin{figure}[htbp]
\centerline{\includegraphics[scale=.35]{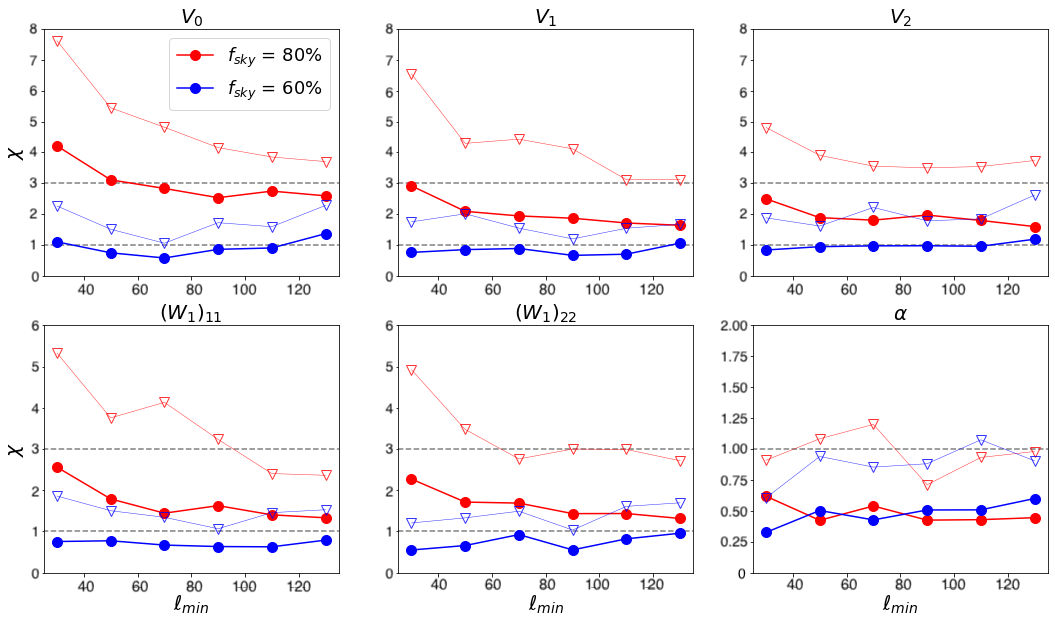}}
\caption{ The three MFs (top), the CMT diagonal terms and $\alpha$ (bottom) deviations as a function of the lower multipole cut, $\ell_{min}$, of the applied band-pass filter. The dots and the triangles represent respectively the average and 95\% percentile values computed over all threshold values.}
\label{fig:MTls}
\end{figure}

 In figure~\ref{fig:MTls} we show results for different values of $\ell_{min}$. For the larger sky fraction ($f_{sky}=80\%$), we find that the deviation decreases as $\ell_{min}$ increases, however, the MF deviation for some thresholds remains significantly high (${>}3\sigma$), even at the smallest scales considered 
 ($\ell_{min} = 130$). For the smaller sky fraction ($f_{sky}=60\%$), the deviations remain almost constant with $\ell_{min}$, indicating consistency between the data and simulations for all thresholds and multipoles at the $3\sigma$ level. 

Referring back to figure~\ref{fig:maskFilMaps}, we see that the areas where several filaments (IIIn, IIIs, IV, VII, X, XVI, XVII) have been detected are masked by the 60\% mask and not the 80\% mask. This suggests that the contribution of the sky regions where these large complex structures are present affects the emission on scales smaller than ${\sim}6^\circ$.

The non-Gaussian deviation decreases for smaller scales, corroborating the hypothesis that at small scales the emission tends to be more Gaussian. However, the effect of bright local structures is still not negligible on a scale of ${\sim}1.5^\circ$. From this analysis we also learn that the non-Gaussian nature is mainly of the kurtosis type, since $V_0$ shows the largest deviations from simulations, as explained in \cite{Rahman21}. These results at 23~GHz are in good agreement with those from the \textit{Planck} 30~GHz polarization maps. More details can be found in appendix~\ref{sec:a03,sub:01}. Moreover, very similar conclusions are obtained for intensity in the analysis of the 408~MHz maps by \cite{Rahman21}.

\section{Non-Gaussian Simulations}
\label{sec:06}

\subsection{A model for non-Gaussian emission}
\label{sec:06,sub:01}
Using MFs and MTs, we have determined clear statistical differences between the real synchrotron polarized emission and an isotropic and Gaussian-distributed model, even at small scales.
In this section, we propose a simple way to generate polarized synchrotron maps which can better resemble the real statistical properties of the emission in the fainter regions of the sky on scales smaller than about $6^\circ$, which corresponds to multipoles $\ell > 30$. 

From our statistical analysis, it is impossible to disentangle non-Gaussianity from anisotropy. That is, we are unable to determine whether the measured non-Gaussianity is an intrinsic feature of the synchrotron emission, or is due to an underlying anisotropic pattern of emission on small scales. However, even if we assume the latter case, because of the lack of information about the small scale distribution of the emission, we need a mechanism to simulate the non-Gaussianity.

We generate the simulations as follows. We start with three complex vectors, $\zeta^T$, $\zeta^E$, $\zeta^B$, generated from a Gaussian random distribution with zero mean and unit variance. The $\zeta^E$ and $\zeta^B$ coefficients are then transformed according to the \textit{sinh-arcsinh} transformation \cite{Pewsey2009}
\begin{equation}
    \Tilde{\zeta^E} = \sinh( \delta \ \textrm{arcsinh}(\zeta^E) - \epsilon )
\label{eq:Htr1}
\end{equation}
\begin{equation}
    \Tilde{\zeta^B} = \sinh( \delta \ \textrm{arcsinh}(\zeta^B) - \epsilon )
\label{eq:Htr2}
\end{equation}
where $\zeta^E$ and $\zeta^B$ are the initial Gaussian-distributed vectors and $\Tilde{\zeta^E}$ and $\Tilde{\zeta^B}$ are the transformed ones. The use of this transformation is motivated by the fact that it allows one to control the level of non-Gaussianities with two parameters, $\epsilon$ and $\delta$, whose effects have a simple statistical interpretation. The parameter $\delta$ introduces symmetrically both positive and negative tails to the statistical distribution, which increases the excess kurtosis. The parameter $\epsilon$ controls the level of skewness.
We find that the Gaussian case is correctly recovered with $(\epsilon, \delta )\rightarrow (0,1)$, when performing simple consistency tests. 

We assume, for the sake of simplicity, that the level of non-Gaussianity is the same for $E$ and $B$ in the part uncorrelated to $T$. This explains why we use the same parameters $(\epsilon, \delta )$ for both transformations in equations~\ref{eq:Htr1}-\ref{eq:Htr2} and we do not transform $\zeta^T$. We are aware that in a more realistic context this could not be the case, but this goes beyond the scope of this analysis. The values used for $(\epsilon, \delta )$ are discussed in section \ref{sec:06,sub:02}. We then generate the spherical harmonic coefficients as follows
\begin{equation}
    a_j^T = \sqrt{C_\ell^{TT}} \zeta^T
\end{equation}
\begin{equation}
    \Tilde{a}_j^E = \frac{C_\ell^{TE}}{\sqrt{C_\ell^{TT}}} \zeta^T + \sqrt{C_\ell^{EE} - \frac{(C_\ell^{TE})^2}{C_\ell^{TT}}} \Tilde{\zeta^E}
\end{equation}
\begin{equation}
    \Tilde{a}_j^B = \frac{C_\ell^{TB}}{\sqrt{C_\ell^{TT}}} \zeta^T + \sqrt{C_\ell^{BB} - \frac{(C_\ell^{TB})^2}{C_\ell^{TT}}} \Tilde{\zeta^B}
\end{equation}
where the index $j$ refers to the $(\ell, m)$ pair. The $C_\ell$ values are computed from the \textit{WMAP} yearly maps as described in section~\ref{sec:05,sub:03}. Note that after rescaling, the $\Tilde{a}_{lm}$-distributions are different from the $\Tilde{\zeta}$-distributions, but still non-Gaussian. We generate the ($I, \ \Tilde{Q}, \ \Tilde{U})$ maps, which represent our small-scale template, from the $\Tilde{a}_{lm}$ using the \texttt{alm2map} routine of \texttt{HEALPix}. 

\subsection{Model for anisotropic modulation of the emission}
Besides the non-Gaussian nature, we have shown that at high multipoles the synchrotron emission is not isotropic. We simply assume that this is due only to a modulation caused by the large bright structures which are also visible at large scales. To simulate the effect, we divide the polarization \textit{WMAP} map into 3 patches ($p_i$) according to the application of thresholds on $P$. We divide the unmasked 80\% of the sky into two patches of 20\%, $p_1$: 80-60\%, $p_2$: 60-40\%, and one of 40\%, $p_3$: 40-0\%, according to the sky brightness. For each patch, we smooth the boundaries with a $5^\circ$ Gaussian and compute the polarization spectra $C_\ell^P(p_i)$. Then we define a spatially varying normalization factor
\begin{equation}
    N_i = \sqrt{ \left<\frac{C_\ell^P(p_i)}{C_\ell^P} \right> }
\end{equation}
where the $C_\ell^P$ is computed from 80\% of the sky and the average is computed over the multipoles $\ell \in [30,180]$. $N$ is then smoothed with a $10^\circ$ beam. As shown in figure~\ref{fig:NGin} (top left), this method naturally introduces the effect at low-scales of most of the filamentary structures presented in section~\ref{sec:04}. We multiply the non-Gaussian $\Tilde{Q}$ and $\Tilde{U}$ maps, computed in the previous section, by the normalization factor  
\begin{equation}
    Q = N\Tilde{Q}, \quad  U = N\Tilde{U}.
\end{equation}
The resulting $Q$ and $U$ maps form the final set of anisotropic and non-Gaussian simulations.

To our small-scale model, we add a large-scale template generated directly from the \textit{WMAP} data. In order to match the maps, we smooth the small-scale and the large-scale templates with respectively the functions $W(\ell)$ and $(1-W(\ell))$, where
\begin{equation}
    W(\ell) = \frac{1}{2}\left( 1 - \tanh\left(\frac{\ell - \ell_0}{\Delta \ell_0}\right) \right),
\end{equation}
with $\ell_0 = 20$ and $\Delta \ell = 5$, chosen such that at the scales of interest in this work, that is smaller than $6^\circ$, the simulations are mainly driven by our small-scale model. Finally, in order to compare simulations with data, we add noise and compute the debiased polarized intensity with the MAS estimator. 

\subsection{Tuning and Results}
\label{sec:06,sub:02}
We have described above the method we use to generate polarized synchrotron simulations on scales below $6^\circ$. The level of non-Gaussianity of the simulations is controlled by the two parameters $(\epsilon, \delta )$ in equation~\ref{eq:Htr1}-\ref{eq:Htr2}, which are related respectively to the skewness $S$ and kurtosis $K$ of the maps. 

\begin{figure}[htbp]
\centerline{\includegraphics[scale=.5]{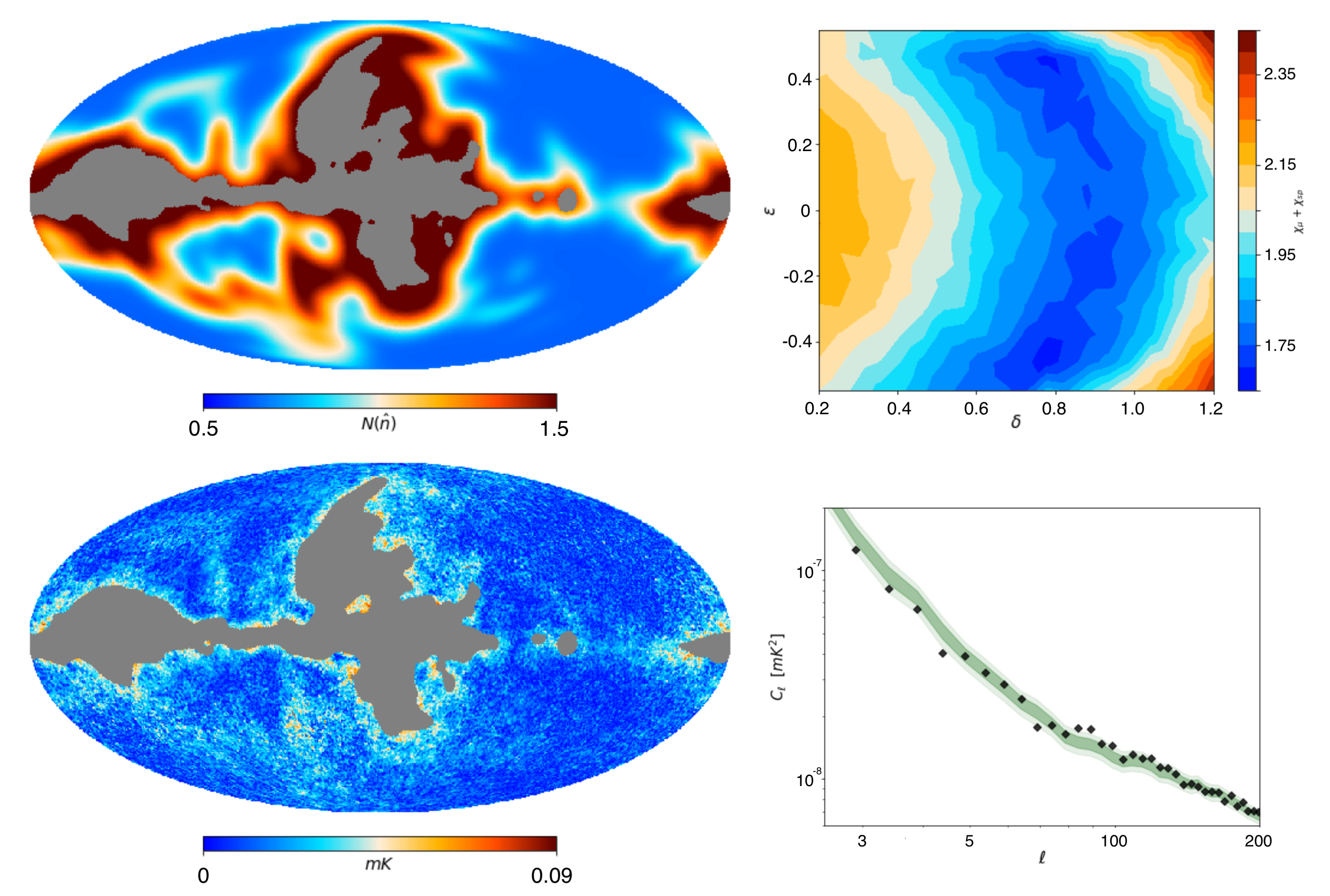}}
\caption{Top Left: Spatially varying normalization factor used to modulate the simulations in real space. Top Right: $\chi_{sp}+\chi_{\mu}$ as function of the parameters $\epsilon$ and $\delta$, on logarithmic scale. Bottom Left: Simulation of the \textit{WMAP} polarized intensity map. Bottom Right: Total polarization power spectrum of the \textit{WMAP} K-band (black dots) compared to the 1$\sigma$ (dark green) and 2$\sigma$ (light green) intervals obtained from the variance of the simulations.}
\label{fig:NGin}
\end{figure}

Similarly to equation~\ref{chiMF}, we define the quantities which measure the deviation between data and simulations as
\begin{equation}
    \chi_{\mu} = \sum_{k=2}^4 \frac{\Delta \mu_k}{\sigma_{\mu_k}}, \quad  \Delta \mu_k = |\mu_k^{data} - \mu_k^{sim}|
\label{chiSkuKor}
\end{equation}
where $\mu_2 = \sigma_P = \sqrt{\langle(p - \langle p \rangle)^2 \rangle}$ is the standard deviation, $\mu_3 = S = \langle(p - \langle p \rangle)^3/\sigma_P^3 \rangle$ is the skewness and $\mu_4 = K = \langle(p - \langle p \rangle)^4/\sigma_P^4\rangle$ is the kurtosis. The values $\mu_k^{sim}$ and $\sigma_{\mu_k}$ are computed taking respectively the average and standard deviation over the simulations. Moreover, we define a variable which quantifies the spectral deviations as
\begin{equation}
    \chi_{sp} = \sum_{\ell=30}^{180}\frac{\Delta C_\ell}{\sigma_{\ell}}, \quad  \Delta C_\ell = |C_\ell^{data} - C_\ell^{sim}|
\label{chiCl}
\end{equation}
where $C_\ell^{sim}$ and $\sigma_{\ell}$ are computed taking respectively the average and standard deviation over the simulations. Note that the $C_\ell$ are binned with a range $\Delta \ell = 10$.
We vary $\epsilon$ and $\delta$ over the ranges $(-1, \ 1)$ and $(0.3, \ 1.5)$, respectively. For each pair of values, we generate 100 total polarization simulations, smoothing them with the band-pass filter ($\ell_{min} = 30$, $\ell_{min} = 180$).  We find that the quantity $\chi_{\mu} + \chi_{sp}$ is minimised for $\epsilon=-0.46$ and $\delta=0.78$, as shown in figure~\ref{fig:NGin} (top right). These define the values to generate a set of reference simulations. An example of a simulated sky, and the comparison between the simulated and real power spectra are shown in figure~\ref{fig:NGin} (bottom). 

\begin{figure}[htbp]
\centerline{\includegraphics[scale=.72]{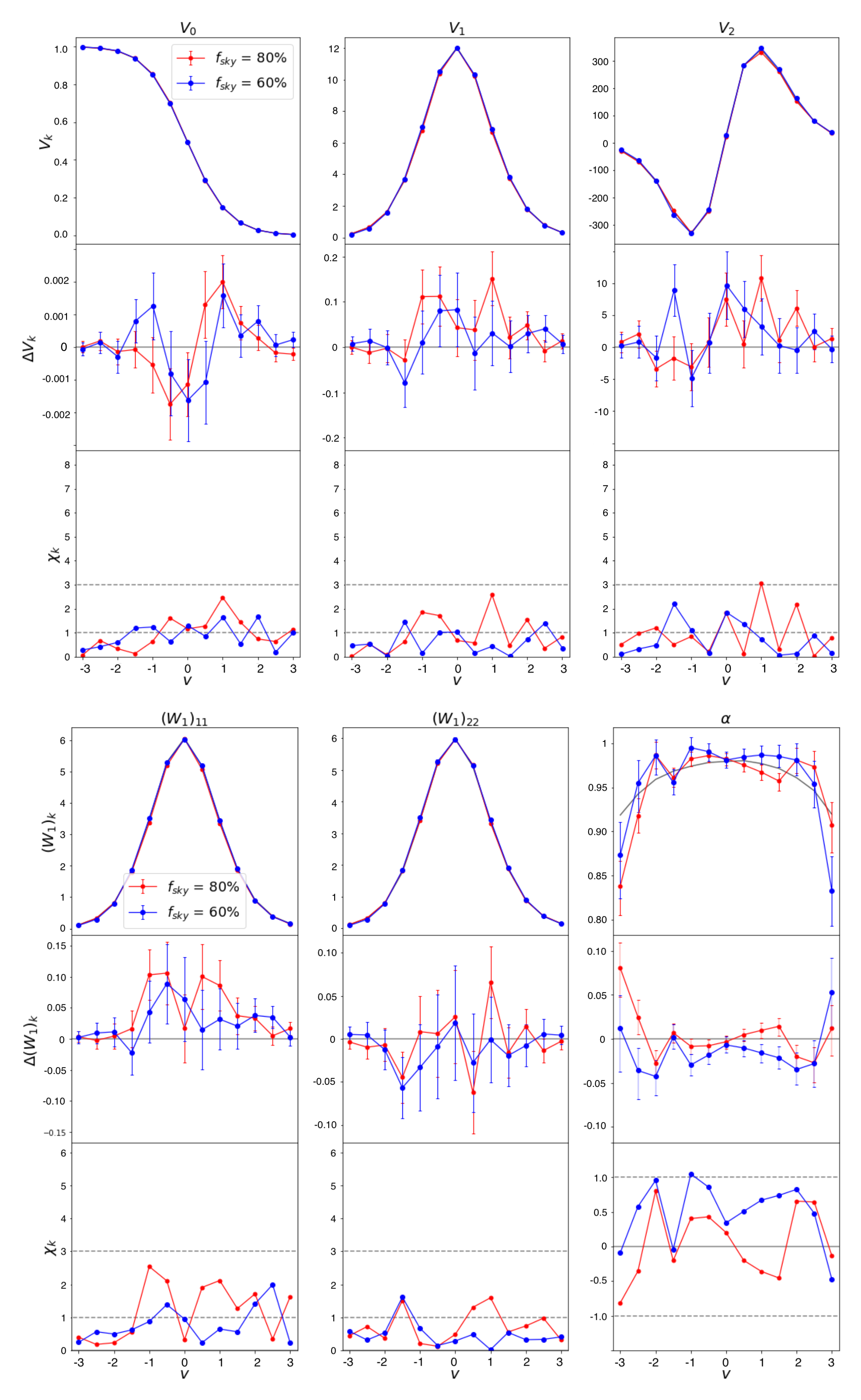}}
\caption{Upper panel: MFs (top), the difference between data and \textit{non-Gaussian} simulations (middle) and $\chi_k$ (bottom) as a function of threshold. Lower panel:  CMT diagonal terms and $\alpha$ (top), difference between data and \textit{non-Gaussian} simulations (middle) and $\chi_k$ (bottom) as a function of threshold. The maps are previously filter with a band-pass ($\ell = 30-180$), error bars are computed from the standard deviation of the simulations. For the bottom panels, we use the same axis ranges as in figure~\ref{fig:MT30} to allow a direct comparison.}
\label{fig:MT30NG}
\end{figure}

\begin{figure}[htbp]
\centerline{\includegraphics[scale=.35]{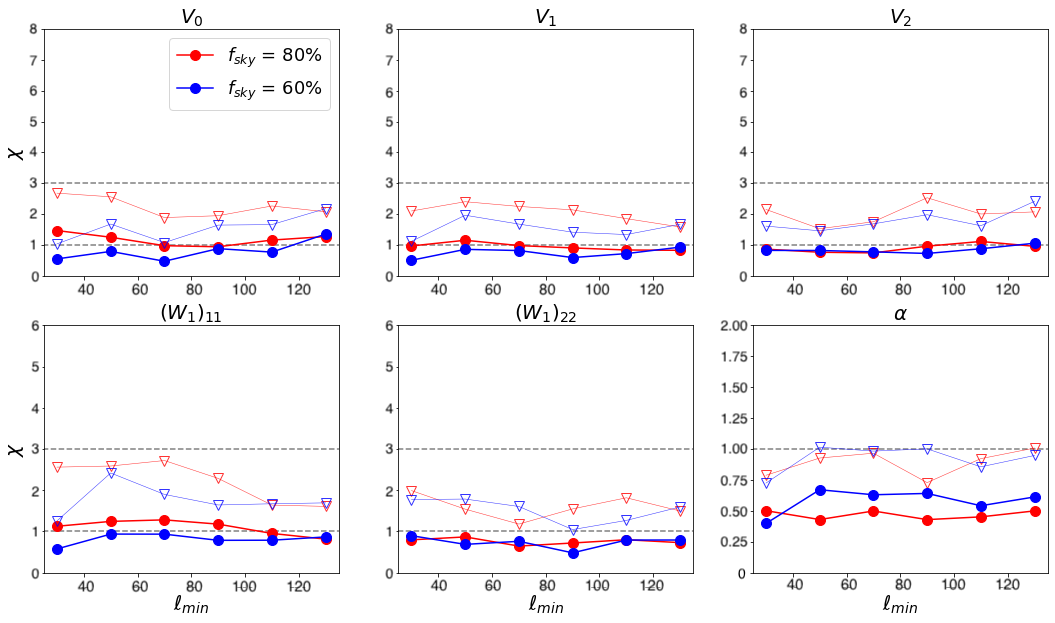}}
\caption{ The three MFs (top), the CMT diagonal terms and $\alpha$ (bottom) deviations from the \textit{non-Gaussian} simulations as function of the lower multipole cut, $\ell_{min}$, of the applied band-pass filter. The dots and the triangles represent respectively the average and 95\% percentile computed over all threshold values. For comparison purposes, we use the same axes as in figure~\ref{fig:MTls}.}
\label{fig:MTlsNG}
\end{figure}

The deviation in the MF and CMT measures between data and simulations, when the band-pass ($\ell_{min} = 30$, $\ell_{min} = 180$) filter is applied, are shown in figure~\ref{fig:MT30NG}. In figure~\ref{fig:MTlsNG}, we show the deviations for different $\ell_{min}$. The simulations agree with the data at the $3\sigma$ level for both the 80\% and 60\% masks. This suggests that the non-Gaussianity introduced with the \textit{sinh-arcsinh} transformation (equations~\ref{eq:Htr1}-\ref{eq:Htr2}) is more relevant in the bright regions, instead is mitigated in the more faint sky emission. As noted in section~\ref{sec:06,sub:01}, we do not know if the non-Gaussianity we introduce is related to the intrinsic nature of the synchrotron emission, or due to an underlying anisotropic emission at small scales. However, the mechanism to produce this non-Gaussianity provides a simple method to tune our simulations by means of two simple parameters ($\epsilon, \delta$). Overall, the non-Gaussian and anisotropic simulations seem to reproduce the statistical properties of the polarized synchrotron emission well. 

In appendix~\ref{sec:a03,sub:02}, we test how our method performs simulating the synchrotron emission at 30~GHz, as observed by \textit{Planck}. We keep the parameters $\epsilon$, $\delta$ and $N$ as fitted from \textit{WMAP}. We observe that the method performs well at small scales (${>}2.5^\circ$), but under-performs for larger scales. 

The goal of this section is to show a simple data-driven way to generate more realistic simulations. There are several assumptions and limitations which are negligible at the current data sensitivity, but will not be the case for future experiments. For example, we assume that the $E$ and $B$ modes are equally non-Gaussian, even if we do not have evidence of it. We do not take into account the correlation of the non-Gaussianity between intensity and polarization. We consider $\epsilon$ and $\delta$ as fixed values, but in a more realistic context, they could vary with scale and space. Moreover, a frequency dependence in our model ($\epsilon$, $\delta$, $N$), for example due to a possible decorrelation between frequencies as observed in the dust emission \cite{planckDust}, could not be excluded. Such effects could explain the poor performance at 30~GHz for the large scales.

\section{Summary and Conclusions}
\label{sec:07}
In the analysis we have covered two important aspects of the polarized synchrotron emission: the presence of large filamentary structures outside the Galactic plane, and its statistical proprieties at small scales. The analysis has been performed on the debiased polarized amplitude maps at 23 and 30~GHz as observed respectively by \textit{WMAP} and \textit{Planck}.
We developed a filament finder routine based on a friend-of-friend recursive algorithm, which detects elongated coherent emission in the sky. The method has been tested with foreground simulations including a toy model of filamentary structure. We identify 19 filaments which are detected, at least in part, in both \textit{Planck} and \textit{WMAP}. Some of the filaments have been already reported in the literature as observed in radio sky or in previous \textit{WMAP} analysis. Five of them are reported for the first time in this work. We analysed some properties of the detected filaments.
\begin{itemize}
\item We compute the polarization fraction as the ratio of the \textit{WMAP} and \textit{Planck} data maps with respect to a \texttt{Commander} intensity template rescaled with a spectral index $\beta_I= -3$. 
Typically, we find that the polarization fractions of the filaments are larger than for the areas outside the filaments, excluding the Galactic plane. For two filaments, both located in the NPS, we find values of about 30\%. 
\item We study the polarization spectral indices of the filaments from the $Q$, $U$ maps. We find consistent spectral indices of about -3.1 inside and outside the filaments. However, the $\beta$ values span a very broad range, from -3.59 to -2.17. 
\item Applying the filament finder algorithm to the $P_E$ and $P_B$ maps, we find that most of the filaments, especially in the NPS, are clearly visible in $E$, but not in $B$. 
\item Applying the finder method to the Haslam map, we observe that for some bright filaments in polarization we do not detect bright counterparts in intensity.
\end{itemize}

Using Minkowski functionals and tensors, we have analysed the non-Gaussianity and statistical isotropy of the polarised \textit{WMAP} and \textit{Planck} maps. We focused on the the faintest 80\% and 60\% of the sky. We compared the results obtained from data with a set of Gaussian and isotropic simulations. We summarize our findings. 
\begin{itemize}
\item Analysing the 80\% sky fraction, we see large deviations ($>3\sigma$) from Gaussianity and isotropy at $6^\circ$ scale. The deviations decrease towards smaller scales, even if they remain significantly high down to $1.5^\circ$.
\item Analysing the 60\% sky fraction, we obtain consistent results between data and simulations for all the considered thresholds and multipole ranges
at the $3\sigma$ level.
\item These results suggest that the large filaments are the main source of non-Gaussianity, even at small scales. When those filamentary structures are masked, the Gaussian and isotropic simulations resemble well the diffuse emission at the \textit{WMAP} and \textit{Planck} resolution.
\end{itemize}

Finally, we present a simple data-driven method used to generate non-Gaussian and anisotropic simulations. We generate non-Gaussian harmonic coefficients by mean of a simple transformation. We account for the anisotropies with a normalization template that resemble the diffuse filamentary structures. The simulations are then fitted in order to match the spectral and statistical properties of the 80\% sky coverage of the data maps.

\acknowledgments
The authors would like to thank the Spanish Agencia Estatal de Investigaci\'on (AEI, MICIU) for the financial support provided under the projects with references PID2019-110610RB-C21, ESP2017-83921-C2-1-R and AYA2017-90675-REDC, co-funded with EU FEDER funds, and acknowledge  support from Universidad de Cantabria and Consejer{\'\i}a de Universidades,  Igualdad,  Cultura  y  Deporte  del  Gobierno de Cantabria via the “Instrumentaci{\'o}n y ciencia de datos para sondear la naturaleza del universo” project as well as from the  Unidad de Excelencia Mar{\'\i}a de Maeztu (MDM-2017-0765). FAM is supported by a fellowship funded by the Unidad de Excelencia María de Maeztu. The authors also thank M. López-Caniego for providing point source masks for CMB data. 
We acknowledge Santander Supercomputacion support group at the University of Cantabria who provided access to the supercomputer Altamira Supercomputer at the Institute of Physics of Cantabria (IFCA-CSIC), member of the Spanish Supercomputing Network, for performing simulations/analyses. This research used resources of the National Energy Research Scientific Computing Center (NERSC), a U.S. Department of Energy Office of Science User Facility located at Lawrence Berkeley National Laboratory, operated under Contract No. DE-AC02-05CH11231.
Some of the presented results are based on observations obtained with \textit{Planck}\footnote{http://www.esa.int/Planck}, an ESA science mission with instruments and contributions directly funded by ESA Member States, NASA, and Canada.
We also acknowledge the Legacy Archive for Microwave Background Data Analysis (LAMBDA). Support for LAMBDA is provided by the NASA oﬃce of Space Science.
Some of the results in this work have been derived using the \texttt{HEALPix} \cite{Healpix,healpy1}, \texttt{NaMaster} \cite{namaster}, \texttt{PySM} \cite{PySM}, \texttt{Matplotlib} \cite{pltpy}, \texttt{NumPy} \cite{NumPy} and \texttt{SciPy} \cite{SciPy} Python packages. 

\appendix

\section{Power Spectra}
\label{sec:a01}

In this section, we give a very brief review of the statistical quantities we use in this work, motivated by standard cosmological practises. CMB experiments usually produce data in the form of three pixelized maps, $T$ for intensity and $Q$ and $U$ Stokes parameters for polarization. 
On the sky, these fields are usually expanded in terms of spherical harmonics
\begin{equation}
    T = \sum_{\ell m} a_{\ell m}Y_{\ell m}
\end{equation}
\begin{equation}
    (Q \pm iU) = \sum_{\ell m} a_{\pm2,\ell m \ } {}_{\pm2}Y_{\ell m}
\end{equation}
where $Y_{\ell m}$ and ${}_{\pm2}Y_{\ell m}$ are respectively the standard and tensor (spin-2) spherical harmonics on a 2-sphere. The quantities $a_{\ell m}$ are the so-called spherical harmonic coefficients. Details of the mathematical formalism can be found in \cite{Kamionkowski_1997,Zaldarriaga_1997}. 

If we define the linear combinations
\begin{equation}
    a_{E,\ell m} = -\frac{1}{2}(a_{2,\ell m \ } + a_{-2,\ell m \ }) \quad   a_{B,\ell m} = -\frac{1}{2i}(a_{2,\ell m \ } - a_{-2,\ell m \ })
\end{equation}
we can decompose the polarization emission into two scalar fields, the gradient-like $E$ mode and the curl-like $B$ mode
\begin{equation}
    E = \sum_{\ell m} a_{E,\ell m}Y_{\ell m} \quad  B = \sum_{\ell m} a_{B, \ell m}Y_{\ell m}.
\end{equation}
The harmonic coefficients may be combined into the angular power spectrum
\begin{equation}
    C_\ell^{XY} = \frac{1}{2\ell+1}\sum_{m} \langle a^*_{X, \ell m}a_{Y, \ell m} \rangle, \quad  X,Y= T, E, B
\end{equation}
which represent: the auto-correlations of temperature and polarization modes denoted by $TT$, $EE$, and $BB$, the cross-correlation between temperature and polarization denoted by $TE$ and $TB$, and the cross-correlation between polarization modes denoted by $EB$. For a Gaussian and isotropic field, all the statistical properties are captured by these two-point statistics. 

Experimental observations are affected by the instrumental (beam) response and the pixelization process. The observed maps can then be written as the convolution of the actual sky signal with the instrumental beam ($B$) and the pixel window function ($W$). The latter is a function of the resolution at which the maps are produced.
In harmonic space, it implies that
\begin{eqnarray}
    a^{obs}_{\ell m} &=& a_{\ell m} B_\ell \ W_\ell \\
        C_\ell^{obs} &=& C_\ell \ B_\ell^2 \ W_\ell^2
\end{eqnarray}
where $B_\ell$ and $W_\ell$ are respectively the harmonic transformations of the instrumental beam and the pixel window function. When analyzing maps produced by different experiments at resolution different from the one in which the original maps are produced, as discussed in this work, it is appropriate to smooth the maps to a common resolution. This can be achieved in harmonic space by
\begin{equation}
    a^{out}_{\ell m} = a^{in}_{\ell m} \ \frac{B^{out}_\ell}{B^{in}_\ell} \ \frac{W^{out}_\ell}{W^{in}_\ell}
\end{equation}
where $B^{in}_\ell$ and $B^{out}_\ell$ are respectively the instrumental and the required beams, $W^{out}_\ell$ and $W^{in}_\ell$ are the pixel window functions at the final and initial resolutions.

\section{Finder Algorithm}
\label{sec:a02}

\subsection{Toy model for filaments}
\label{sec:a02,sub:01}
\begin{figure}[htbp]
\centerline{\includegraphics[scale=.55]{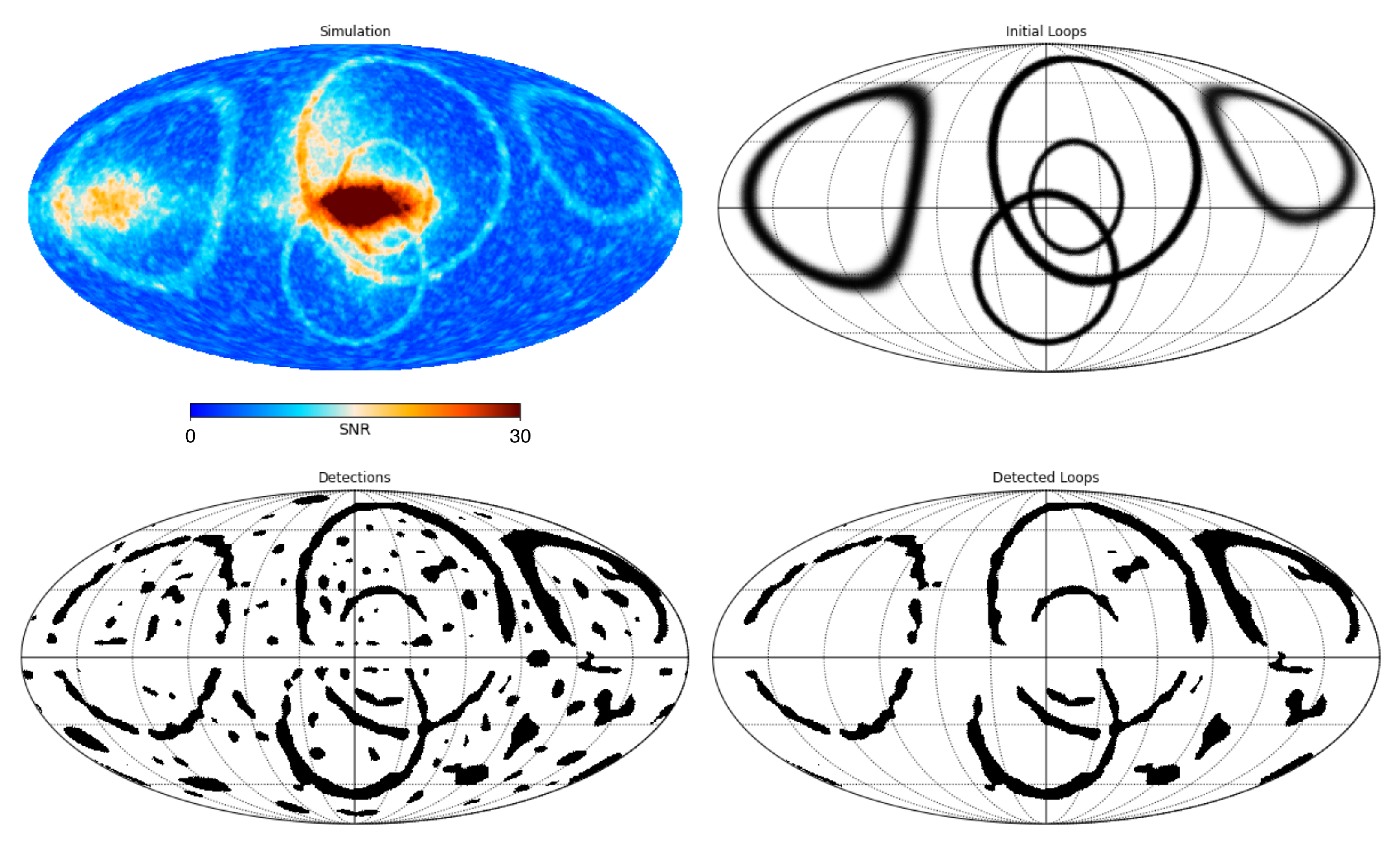}}
\caption{Top Left: Toy filamentary foreground simulation. Top Right: Loop template used in the simulation. Bottom Left: All the detections found with the friends-of-friends recursive algorithm. Bottom Right: final result of the filament finder algorithm.}
\label{fig:toyModel}
\end{figure}

In this section, we test the performance of the filament finder algorithm using foreground simulations including a toy model of filamentary structure. Each simulation is computed as the sum of different independent components
\begin{equation}
    S = S_{Gal} + S_{dif} + S_{Loops} + S_{noise},
\label{eq:ToyModel}
\end{equation}
where:
\begin{itemize}
    \item $S_{Gal}$ is a Galactic plane simulation. The template is generated from the \textit{WMAP} K-band $P$ map, smoothed to a resolution of $5^\circ$ and filtered with a low-pass filter $ f(\ell) = \left[ 1 - \tanh\left( (\ell - 10)/10 \right) \right]/2$. In this way we preserve the Galactic morphology on large scales ($\ell <10$) whilst removing the small scales corresponding to the real filaments.
    \item $S_{dif}$ is a diffuse Gaussian template created with the \texttt{synfast} routine using the power spectra model: $C_\ell \propto (\ell/80)^{-2.9}$ \cite{Martire_2022}. The simulated map is filtered with a high-pass filter $f(\ell) = \left[ 1 - \tanh\left( (\ell + 10)/10 \right) \right]/2$, which only allows multipoles $\ell>10$.
    \item $S_{Loops}$ is a template where different loops are projected onto the sphere. The loops are based on filaments observed in \textit{WMAP} \cite{Vidal_2015}. Each loop has been generated with a width in the range 2--4$^\circ$. In order to simulate both thin and diffuse filaments, we smooth the loops with a $1.5^\circ$ or a $3^\circ$ Gaussian beam. The loops are shown in figure~\ref{fig:toyModel} (top right).
    \item $S_{noise}$ is a noise simulation with properties estimated from the \textit{WMAP} noise covariance matrices.
\end{itemize}

\begin{figure}[htbp]
\centerline{\includegraphics[scale=.40]{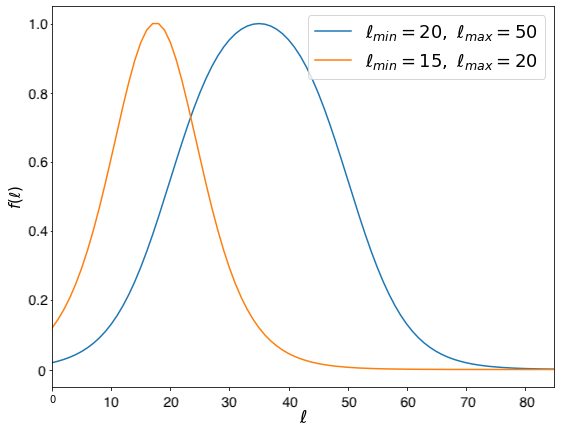}}
\caption{Band-pass filters defined in equation \ref{filter} used for the detection of the bright (blue) and the weak (orange) filaments. }
\label{figfilter}
\end{figure}

The simulation $S$ is produced at $N_{side} = 128$ and a resolution of $1^\circ$, as used for the data. The Galactic and diffuse templates have been re-scaled in order to match the data signal-to-noise ratio. We tested different amplitudes, locations and radii for the loops, although in the following we will only refer to the case including Loops I, III, GCS, VII and XI, at a signal-to-noise ratio of 5, as shown in figure~\ref{fig:toyModel}. 
We filter the maps with the filters in figure~\ref{figfilter} and apply the friends-of-friends algorithm to 100 simulations. Note that each simulation has the same $S_{Gal}$ and $S_{Loop}$, but different realization of $S_{dif} $ and $S_{noise}$.
Figure~\ref{fig:toyModel} presents an example of a simulation (top left), and the corresponding detected structures before (bottom left) and after (bottom right) the minimal length criteria is applied, as described in section~\ref{sec:03,sub:01}.

For each simulation, we recover on average 71.0\% ($\pm 1.4\%$) of the original filaments. However, we also assign a detection of filamentary structure to around 7.2\% ($\pm 0.7\%$) of the sky which is not associated with any input loops. From figure~\ref{fig:toyModel}, we observe that the filament finder mostly fails to detect parts of filaments close or tangential to the Galactic plane, where the strong Galactic emission dominates. We also point out that the detection can fail in those areas where two or more loops overlap, because the orientation angle in those pixels is the result of the average over different loops.
The detections that are not associated with any input loop mainly arise in the regions with the lowest signal-to-noise ratio, suggesting that the noise is the cause. However, we note that it is possible to identify most of these spurious detections by comparing two simulations with different noise realisations. In practice, in our main analysis with real data, this is achieved by comparing the results of two independent maps, from \textit{WMAP} and \textit{Planck}, which allows us to reduce the number of spurious detections. 

Note that the quantitative results presented in this appendix are obtained with reference to the \textit{WMAP} data. However, all qualitative considerations also apply to the \textit{Planck} data. Possible differences in the performance of the algorithm are mainly attributable to the fact that the difference of the filament brightness to the diffuse background is greater in the \textit{WMAP} map than in the \textit{Planck} map. In addition, the different distributions of noise for the two experiments could also have an impact on the performance of the algorithm, although we expect it to be subdominant.

\subsection{Minimal length criteria}
\label{sec:a02,sub:02}
The filament finder method presented in section~\ref{sec:03,sub:01} is a simple friends-of-friends recursive algorithm based on the properties of single pixels. When a group of coherent and bright pixels is identified, it is not obvious if it is part of a filamentary structure or not. Considering the positive nature of the polarization intensity, regions where the noise is strong can confuse the detection. Moreover, the synchrotron diffuse background can also have a detrimental effect. In order to reduce spurious detections, we reject structures where the maximum pixel-pair angular distance is smaller than a threshold value $L_{th}$. 

We find the best value for $L_{th}$ by analysing simulations that do not contain loop structures. As in section~\ref{sec:a02,sub:01}, we generate 100 diffuse synchrotron simulations from $S = S_{Gal} + S_{dif} + S_{noise}$, where the single components are described in the previous section. Note that in this case the simulations do not include the $S_{Loops}$ term. 

\begin{figure}[htbp]
\centerline{\includegraphics[scale=.55]{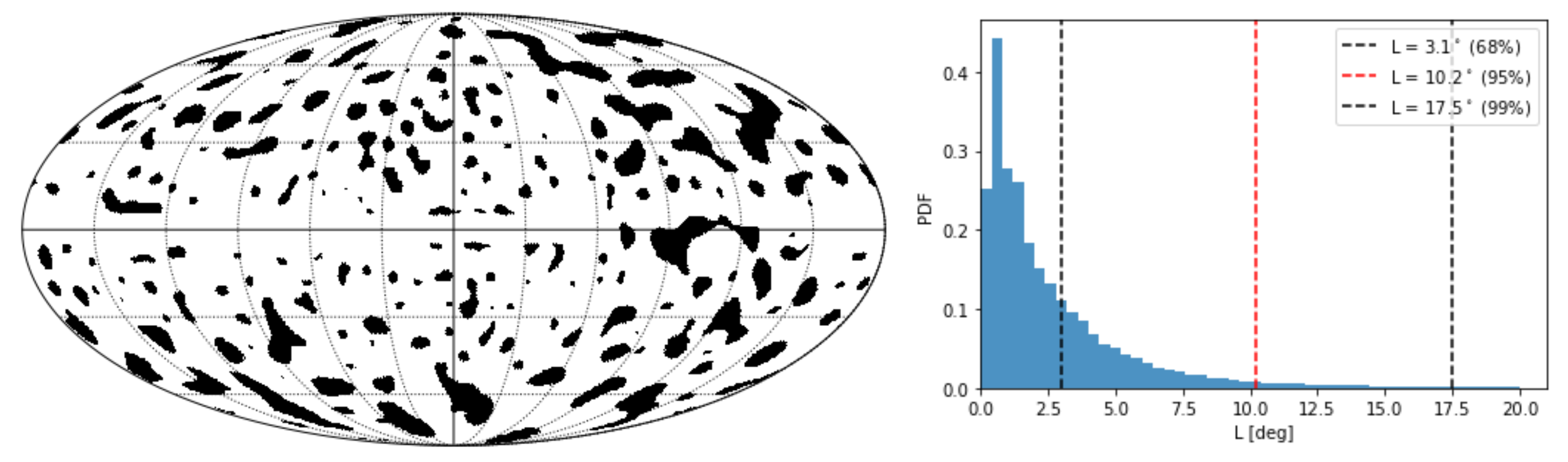}}
\caption{Left: All the detections found with the friends-of-friends recursive algorithm from a loop-less simulation. Right: Distribution of the maximum angular lengths obtained from 100 loop-less simulations.}
\label{fig:toyModelNoFil}
\end{figure}

We apply the finder algorithm to each simulation which now can only detect spurious signals due to noise and the diffuse emission. Figure~\ref{fig:toyModelNoFil} shows an example of a detection (left) and the distribution of the lengths determined from the simulations (right). 
We find that 68\% of detections have a length smaller than roughly $3.1^\circ$, 95\% smaller than $10.2^\circ$ and 99\% smaller than $17.5^\circ$. From this result, we pick the threshold value $L_{th} = 10^\circ$. Note that this estimate holds for the pessimistic scenario of a loop-less foreground. In a more realistic case, i.e. including bright filaments, the algorithm would rely on a larger $P_{th}$, so a part of the noise detection would not exceed the threshold, and we would get less spurious detections.

\section{\textit{Planck} Statistical Properties}
\label{sec:a03}

\subsection{Gaussian Simulations}
\label{sec:a03,sub:01}
In the main text, we analyse the statistical properties of the \textit{WMAP} K-band maps. The choice is motivated by the fact that at 23~GHz the synchrotron emission is much stronger than at the 30~GHz \textit{Planck} frequency channel. However, as a consistency check, in this section we present the results obtained when analysing the \textit{Planck} data. We use the same masks and filters presented in section~\ref{sec:05,sub:02}. 

In order to quantify the non-Gaussianity and anisotropy of the synchrotron emission observed by \textit{Planck}, we need to compare data with a set of suitable simulations. We compute pseudo-spectra in the unmasked regions cross-correlating A/B split maps\footnote{The PR4 provides A/B splits for data maps and simulations \cite{Npipe}. For the 30~GHz frequency channel, the A and B subsets are obtained respectively combining maps from years 1 and 3, and years 2 and 4.}. From the spectra, we generate 600 Gaussian and isotropic simulations of $Q$ and $U$, add noise, then compute the debiased polarized intensity. 

\begin{figure}[htbp]
\centerline{\includegraphics[scale=.35]{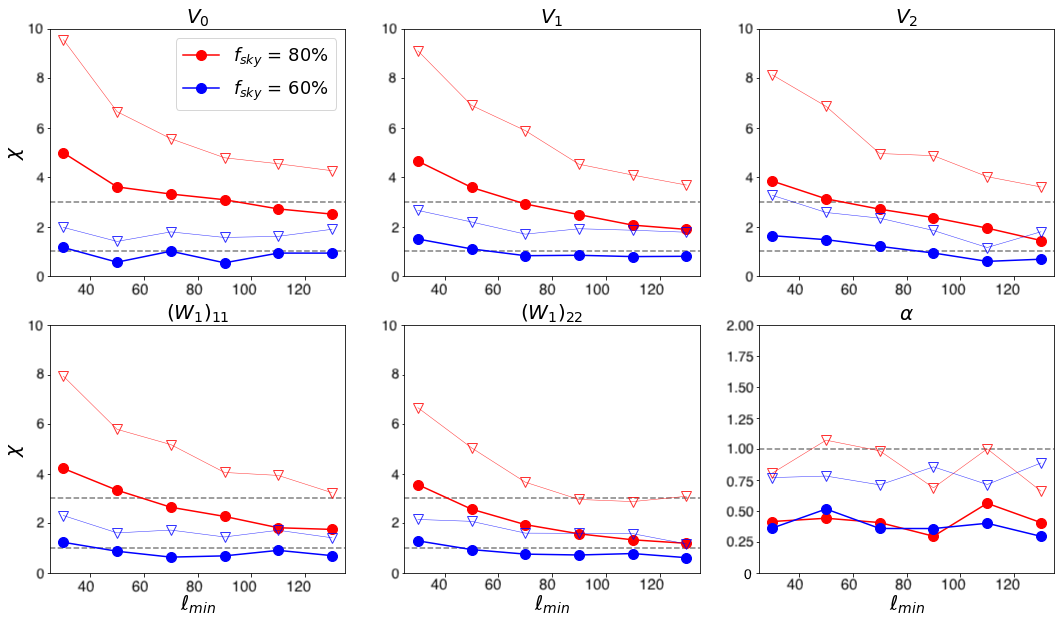}}
\caption{Top: The three MFs, and bottom: the CMT diagonal terms and the $\alpha$ deviations from the Gaussian simulations, computed with \textit{Planck} data, as a function of the low multipole cut $\ell_{min}$ from the applied band-pass filter. The dots and the triangles represent respectively the average and 95\% percentile values computed over all threshold values.}
\label{fig:MTlsPl}
\end{figure}

Results are shown in figure~\ref{fig:MTlsPl}. For the larger sky fraction ($f_{sky}=80\%$), we find that for the cases with $\ell_{min}<80$, even averaging over the thresholds, the deviation exceeds $3\sigma$. The deviation decreases when $\ell_{min}$ increases, however, for all the quantities (except $(W_1)_{22}$), some thresholds remain significantly higher than $3\sigma$. For the smaller sky fraction ($f_{sky}=60\%$), we generally find consistency between the data and simulations. 
These results are in substantial agreement with those determined with \textit{WMAP} at 23~GHz, corroborating the discussion in section~\ref{sec:05,sub:04}.

\subsection{Non-Gaussian Simulations}
\label{sec:a03,sub:02}
\begin{figure}[htbp]
\centerline{\includegraphics[scale=.35]{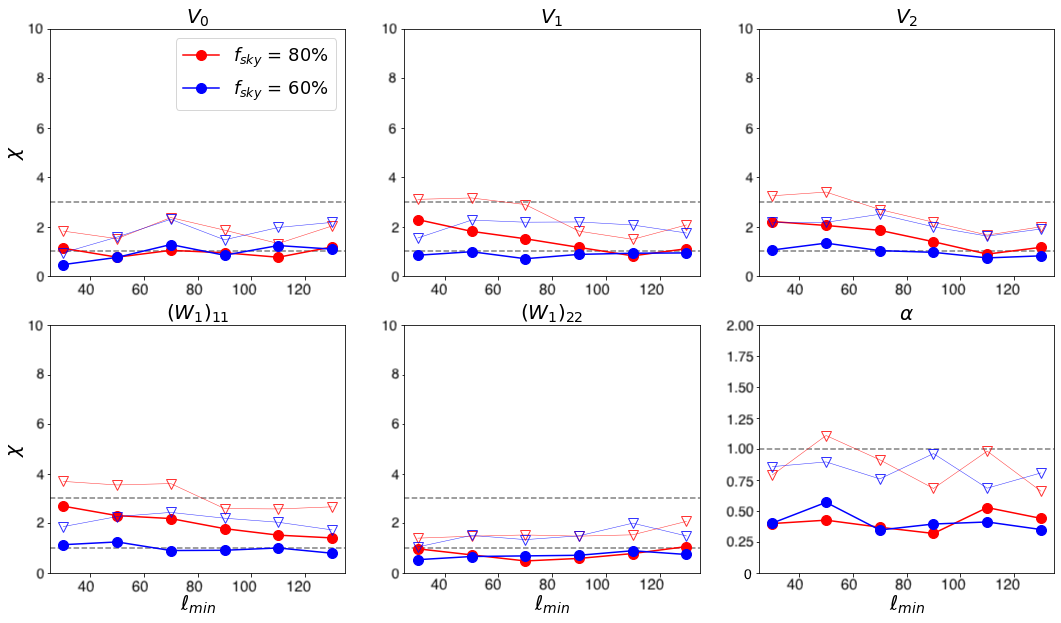}}
\caption{Top: The three MFs, and bottom: the CMT diagonal terms and $\alpha$ deviations from the \textit{non-Gaussian} simulations, computed with \textit{Planck} data, as a function of the low multipole cut $\ell_{min}$ from the applied band-pass filter. The dots and the triangles represent respectively the average and 95\% percentile computed over all threshold values. For comparison purposes, we use the same ranges as in figure~\ref{fig:MTlsPl}.}
\label{fig:MTlsPlNG}
\end{figure}

In section~\ref{sec:06}, we present a data-driven method to simulate the polarized synchrotron emission at 23~GHz. In this section, we show how the model performs in reproducing the emission at 30~GHz. We use the same spatially varying normalization factor (see figure~\ref{fig:NGin}) and $(\epsilon,\ \delta)$ parameters to introduce anisotropies and non-Guassianities as for \textit{WMAP}. The results are shown in figure~\ref{fig:MTlsPlNG}. 

The simulations agree with the data at the $3\sigma$ level for the 80\% mask for those cases with $\ell_{min}>70$, and for all the considered multipole ranges for the 60\% mask. The model seems to under-perform for the cases with $\ell_{min} \le 70$ when considering the 80\% mask, although the deviations from the data are clearly not so pronounced as when using Gaussian simulations. Considering that the largest deviation comes from $(W_1)_{11}$, it is reasonable to think that we are not correctly taking into account the anisotropy of the field.

It has been shown, even in this work, that the polarization spectral index shows spatial variations, and bright structures at 23~GHz are less detectable at 30~GHz. This suggests that the spatially varying normalization factor computed only from the \textit{WMAP} data, could also depend on frequency. In addition, given that the $V_1$ and $V_2$ values computed with simulations deviate from the data when considering multipoles $\ell<60$, we can not exclude the possibility that the non-Gaussianity level could also depend on frequency, which in our model translates into $\epsilon = \epsilon(\nu)$ and $\delta = \delta(\nu)$.

\newpage

\bibliographystyle{JHEP.bst}
\bibliography{main.bib}

\end{document}